\documentclass[12pt]{article}
\usepackage{amsmath,amssymb}
\usepackage{graphicx}
\usepackage{subcaption}
\usepackage{cite}

\def\mydate{3 September 2016}
\def\ignore#1{{}}

\makeatletter
\@addtoreset{equation}{section}
\makeatother

\newcommand{\beeq}{\begin{equation}}
\newcommand{\eneq}{\end{equation}}
\newcommand{\beqn}{\begin{eqnarray}}
\newcommand{\eeqn}{\end{eqnarray}}

\def\mybig{\displaystyle \strut }

\def\dd{\partial}
\def\la{\raise.16ex\hbox{$\langle$}\lower.16ex\hbox{}  }
\def\ra{\raise.16ex\hbox{$\rangle$}\lower.16ex\hbox{} }
\def\go{\rightarrow}

\def\onehalf{ \hbox{$\frac{1}{2}$} }
\def\onethird{ \hbox{$\frac{1}{3}$} }
\def\twothird{ \hbox{$\frac{2}{3}$} }

\def\Tr{{\rm Tr \,}}
\def\tr{{\rm tr \,}}
\def\eff{{\rm eff}}

\def\SM{{\rm SM}}

\def\KK{{\rm KK}}

\def\vect{{\rm vec}}
\def\sp{{\rm sp}}

\def\ep{\epsilon}
\def\psibar{ \psi \kern-.65em\raise.6em\hbox{$-$} }
\def\psibarl{ \psi \kern-.65em\raise.6em\hbox{$-$} \lower.6em\hbox{} }

\def\myfrac#1#2{{\mybig #1\over \mybig #2}}

\newcommand{\longbar}[1]{\raise.6em\hbox{$-$}\kern-0.75em#1 }
\newcommand{\Fbar}{\raise.65em\hbox{$-$}\kern-0.9em F}
\newcommand{\diracslash}[1]{/\kern-0.5em #1}
\def\m{{(m)}}
\def\n{{(n)}}

\begin{document}

\thispagestyle{empty}
{\small \noindent \mydate    \hfill OU-HET 872, KIAS-P15044}
\vspace{3.0cm}

\baselineskip=35pt plus 1pt minus 1pt
\begin{center}
{\LARGE \bf $H \to Z\gamma$ in the gauge-Higgs unification}
\end{center}

\vspace{1.5cm}
\baselineskip=22pt plus 1pt minus 1pt
\begin{center}
{\bf Shuichiro Funatsu$^*$, Hisaki Hatanaka$^\dagger$, Yutaka Hosotani$^*$}
\vskip 5pt
$^*${\small \it Department of Physics, Osaka University, Toyonaka, Osaka 560-0043, Japan} \\
$^\dagger${\small \it Quantum Universe Center, Korea Institute for Advanced Study}\\
{\small \it Seoul 130-722, Republic of Korea}\\
\end{center}

\vskip 2.cm
\baselineskip=20pt plus 1pt minus 1pt
\begin{abstract}
The decay rate of the Higgs decay $H \to Z \gamma$ is
evaluated at the one-loop level in the $SO(5)\times U(1)$ gauge-Higgs
unification. 
Although an infinite number of loops with Kaluza-Klein states contribute to the
decay amplitude, there appears the cancellation among the loops, and
the decay rate is found to be finite and non-zero. It is   found
that the decay rate is well approximated by the decay
rate in the standard model multiplied by $\cos^2\theta_H$, where
$\theta_H$ is the Aharonov-Bohm phase induced by the vacuum
expectation value of an extra-dimensional component of the gauge
field.
\end{abstract}

\newpage

\baselineskip=20pt plus 1pt minus 1pt
\parskip=0pt

\section{Introduction}
The Higgs boson of a mass about 125 GeV has been found at LHC.\cite{Aad:2012tfa}\cite{Chatrchyan:2012ufa}
The signal strength of each decay mode of the Higgs boson has been consistent with 
the standard model (SM).\cite{Aad:2013wqa}\cite{CMS:yva}
Though the decay mode  $H\to Z \gamma$ has not been observed so far, it is 
expected to be seen in the Run 2 at LHC.
The decay rate $\Gamma (H\to Z \gamma)$ has been evaluated in the SM,\cite{Bergstrom:1985hp}
the two Higgs doublet model,\cite{Gunion:1989we} the minimal supersymmetric standard model,\cite{Gunion:1989we}
the universal extra dimension model,\cite{Petriello:2002uu}
and the type-II seesaw model.\cite{Dev:2013ff}

The gauge-Higgs unification (GHU) is one of the scenarios beyond the SM.\cite{Hosotani:1983xw,
Hosotani:1988bm,Davies:1987ei,Davies:1988wt,Hatanaka:1998yp,Burdman:2002se,
Csaki:2002ur,Matsumoto:2014ila,Lim:2014tua}
In GHU the 4D Higgs boson appears as part of the extra-dimensional component of 
the gauge potentials.
When the extra-dimensional space is not simply connected, it is identified with the 4D fluctuation
mode of the Aharonov-Bohm (AB) phase $\theta_H$  along the extra-dimensional space.  
The gauge invariance protects the Higgs boson from acquiring divergent mass corrections.
The Higgs boson mass is generated at  the quantum level,  being finite and  independent of 
the cutoff scales in the theory.
Especially the $SO(5)\times U(1)$ GHU  in the Randall-Sundrum (RS) space-time
is phenomenologically successful.\cite{Agashe:2004rs,Medina:2007hz,Hosotani:2007qw,
Hosotani:2008tx,Hosotani:2009qf}
The Higgs doublet appears in the $SO(5)/SO(4)$ part of the fifth dimensional component 
of the vector potentials with the custodial symmetry.
The model  is consistent with the LHC results for $\theta_H < 0.1$.
The deviation of the decay rate $\Gamma(H\to\gamma\gamma)$ from the SM, for instance, 
is less than 1\%,\cite{Funatsu:2013ni} despite the fact that an infinite number of Kaluza-Klein (KK) modes 
of the $W$ boson and top quark contribute.
$Z'$ events are expected as the excitation of the first KK modes of $\gamma, Z$ and 
the lowest mode of $Z_R$,  the neutral $SU(2)_R$ gauge boson. 
Their masses are almost degenerate, and are estimated to be in the range 
4 to  8.5 TeV for $\theta_H = 0.15 \sim 0.07$.\cite{Funatsu:2014fda}
There also exists a dark matter candidate in the model,
the lowest KK mode of $SO(5)$-spinor fermion called dark fermions.\cite{Funatsu:2014tka} 
Its mass is in the range of 2.3 - 3.1 TeV and the spin-independent scattering cross section 
per nucleon is $\sigma_N\simeq O(10^{-44}) \;\text{cm}^2$.
It may be detected in the 300 live days run of the LUX experiment.

In this paper we focus on the decay mode $H\to Z \gamma$  in the $SO(5) \times U(1)$ GHU.
The decay width of $H\to Z \gamma$ has been evaluated  in the $SU(3)$ GHU model
on flat $M^4 \times (S^1/Z_2)$ by Maru and Okada.\cite{Maru:2013bja}  
They found that it vanishes at the one loop level,
due to the group structure of the $SU(3)$ model.
In the  $SO(5) \times U(1)$ GHU in RS, on the other hand, it is known that the gauge
couplings of the SM particles are almost the same as in the SM so that one expects
that the process $H\to Z \gamma$ occurs.  Furthermore, as in the case of
$H \go \gamma \gamma$, one needs to worry about the contributions coming
from an infinite number of KK modes running in the loops.
The situation in the case of $H\to Z \gamma$ is more involved than that in the case of 
$H\to \gamma \gamma$.
In $H\to \gamma\gamma$, the KK number of virtual particles running the inside loop
is conserved.   In contrast, in the case of $H\to Z \gamma$, the KK number of virtual particles 
may change, as both $H$ and $Z$ have off-diagonal couplings in RS.
This gives rise to an interesting question whether or not the sum of all these contributions
converges.  It seems to require more subtle cancellation mechanism to have a finite
result than in the case of $H\to \gamma\gamma$.
We demonstrate in this paper by direct evaluation that miraculous cancellation
takes place among KK-number-conserving  and  KK-number-violating loops. 
After all, the deviation of the decay width from that in the SM is $O(1)$ \%.

The paper is organized as follows.  
In section 2, the model of the $SO(5) \times U(1)$ GHU is explained.
In section 3, we review the decay rate of the $H\to\gamma\gamma$ 
and evaluate the decay rate of the $H\to Z\gamma$ process  
in the $SO(5) \times U(1)$ GHU. 
In section 4, conclusion  and discussions are given.
In the appendix, we summarize $Z$ and $H$  couplings of various fields 
which are necessary in  calculating the $H \to Z\gamma$ decay rate.

\section{Model}

We consider the $SO(5)\times U(1)$ gauge-Higgs unification 
in the Randall-Sundrum (RS) warped space,\cite{Randall:1999ee} whose metric is given by 
$ds^2=G_{MN}dx^M dx^N=e^{-2\sigma(y)} \eta_{\mu\nu}dx^\mu dx^\nu+dy^2$, 
where $\eta_{\mu\nu} = \text{diag}(-1,1,1,1)$, $\sigma(y)=\sigma(y+2L)=\sigma(-y)$, 
and $\sigma(y)=k|y|$ for $|y| \le L$.
The Planck and TeV branes are located at $y = 0$ and $y = L$, respectively.
The bulk region $0 < y < L$ is anti-de Sitter (AdS) spacetime 
with a cosmological constant $\Lambda = -6k^2$.
The warp factor is $z_L \equiv e^{kL} \gg 1$, and the Kaluza-Klein mass scale is given by 
$m_{KK} = \pi k/(z_L - 1) \sim \pi k z_L^{-1}$.
The model consists of $SO(5)\times U(1)_X$ gauge fields $(A_M , B_M )$, 
quark-lepton multiplets  $\Psi_a$, $SO(5)$-spinor fermions (dark fermions) $\Psi_{F_i}$, 
brane fermions $\hat{\chi}_{\alpha R}$, and brane scalar $\hat \Phi$.\cite{Hosotani:2009qf, Funatsu:2013ni}
The model has been specified in Refs. \cite{Funatsu:2014fda, Funatsu:2014tka}.
The bulk part of the action is given by
\beqn
&&\hskip -1.cm
S_{\text{bulk}}=\int d^5x\sqrt{-G} \, \Bigl[-\tr 
\Bigl(\frac{1}{\;4\;}F^{(A) MN} F^{(A)}_{MN}+\frac{1}{2\xi}
(f^{(A)}_{\text{gf}})^2+\mathcal{L}^{(A)}_{\text{gh}}\Bigr) \cr
\noalign{\kern 10pt}
&&\hskip 2.5 cm
-\Bigl(\frac{1}{\;4\;}F^{(B) MN} F^{(B)}_{MN}+\frac{1}{2\xi}
(f^{(B)}_{\text{gf}})^2+\mathcal{L}^{(B)}_{\text{gh}}\Bigr) \cr
\noalign{\kern 10pt}
&&\hskip 2.5 cm
+\sum_{a} \bar{\Psi}_a\mathcal{D}(c_a)\Psi_a
+ \sum_{i=1}^{n_F} \bar{\Psi}_{F_i} \mathcal{D}(c_{F_i})\Psi_{F_i} \Bigr],\cr
\noalign{\kern 10pt}
&&\hskip -1.cm
\mathcal{D}(c)= \Gamma^A {e_A}^M
\Big(\partial_M+\frac{1}{8}\omega_{MBC}[\Gamma^B,\Gamma^C]
-ig_AA_M -ig_BQ_{X}B_M\Big)-c\sigma'(y) .
\label{action1}
\eeqn
The gauge fixing and ghost terms are denoted as functionals with subscripts gf and gh, respectively. 
$F^{(A)}_{M N}=\partial_M A_N-\partial_N A_M -ig_A\bigl[A_M,A_N \bigr]$, and 
$F^{(B)}_{M N}=\partial_M B_N-\partial_N B_M$. 
The color $SU(3)_C$ gluon fields and their interactions have been suppressed.
The $SO(5)$ gauge fields $A_M$ are decomposed as
\begin{equation}
A_M=\sum^3_{a_L=1}A_M^{a_L}T^{a_L}+\sum^3_{a_R=1}A_M^{a_R}T^{a_R}
+\sum^4_{\hat{a}=1}A_M^{\hat{a}}T^{\hat{a}} ,
\end{equation}
where $T^{a_L, a_R}  (a_L , a_R = 1, 2, 3)$ and $T^{\hat{a}}  (\hat{a} = 1, 2, 3, 4)$ 
are the generators of $SO(4) \simeq SU(2)_L \times SU(2)_R$ and $SO(5)/SO(4)$, respectively.
The quark-lepton multiplets $\Psi_a$ are introduced in the vector representation of $SO(5)$, whereas
$n_F$ dark fermions $\Psi_{F_i}$  in the spinor representation with $Q_X = \onehalf$.
The dimensionless parameter $c$, which gives a bulk kink mass,  plays an important role 
in controlling profiles of  fermion wave functions.  
$\bar{\Psi}=i\Psi^\dagger \Gamma^0$.

The orbifold boundary conditions at $y_0 = 0$ and $y_1 = L$ are given by
\beqn
&&\hskip -1.cm
\begin{pmatrix}A_{\mu}\\A_y\end{pmatrix} (x,y_j-y)=
P_{\vect} \begin{pmatrix}A_{\mu}\\-A_y\end{pmatrix} (x,y_j+y) P_{\vect}^{-1},  \cr \noalign{\kern 10pt}
&&\hskip -1.cm
\begin{pmatrix}B_{\mu}\\B_y\end{pmatrix} (x,y_j-y)=
\begin{pmatrix}B_{\mu}\\-B_y\end{pmatrix}(x,y_j+y), \cr \noalign{\kern 10pt}
&&\hskip -1.cm
\Psi_a(x,y_j-y)=P_{\vect} \Gamma^5 \Psi_a(x,y_j+y),\cr \noalign{\kern 10pt}
&&\hskip -1.cm
\Psi_{F_i} (x, y_j -y)=
\eta_{F_i} (-1)^j P_{\sp} \Gamma^5\Psi_{F_i} (x, y_j + y),  
~~~ \eta_{F_i} = \pm 1 ,  \cr \noalign{\kern 10pt} &&\hskip -1.cm
P_{\vect}=\text{diag} \, ( -1, -1, -1, -1, +1) , \quad 
P_{\sp} =\text{diag}  \, ( +1, +1, -1, -1 ) ,
\label{BC1}    
\eeqn
which reduce the $SO(5)\times U(1)_X$ symmetry   to 
$SO(4)\times U(1)_X \simeq SU(2)_L\times SU(2)_R\times U(1)_X$. 
$SO(4) \times U(1)_X$ symmetry is spontaneously broken to $SU(2)_L \times U(1)_Y$ 
by the brane scalar $\hat \Phi$.

\ignore{
Various orbifold boundary conditions fall into a finite number of equivalence classes of boundary conditions.\cite{Haba:2003ux}
The physical symmetry of the true vacuum in each equivalence class of boundary conditions is dynamically determined at the quantum level by the Hosotani mechanism.
Recently dynamics for selecting boundary conditions have been proposed as well.\cite{Yamamoto:2013oja}
The Hosotani mechanism has been explored and established, not only in perturbation theory, 
but also on the lattice nonperturbatively.\cite{Cossu:2013ora}
}

The 4D Higgs field appears as a zero mode in the $SO(5)/SO(4)$ part of the fifth dimensional 
component of the vector potential $A^{\hat{a}}_y(x,y)$  with custodial symmetry.
Without loss of generality one can set $\langle A^{\hat{a}}_y\rangle 
\propto \delta^{a4}$ when the electroweak symmetry is spontaneously broken. 
The 4D neutral Higgs field $H(x)$ is a fluctuation mode of 
the Wilson line phase $\theta_H$ which is an Aharonov-Bohm phase in the fifth dimension;
\beqn
&&\hskip -1.cm
A^{\hat{4}}_y (x,y)= \big\{\theta_Hf_H+H(x)\big\}u_H(y)+\cdots ~, \cr
\noalign{\kern 10pt}
&&\hskip -1.cm
\exp \Big\{ \frac{i}{2} \theta_H \cdot 2 \sqrt{2}T^{\hat{4}} \Big\} 
= \exp  \bigg\{ ig_A	\int^L_0 dy  \la A_y \ra \bigg\}   ~, \cr
\noalign{\kern 10pt}
&&\hskip -1.cm
f_H=\frac{2}{g_A} \sqrt{\frac{k}{z_L^2-1}}
=\frac{2}{g_w} \sqrt{\frac{k}{L(z_L^2-1)}} ~.
\label{WilsonPhase1}
\eeqn
Here the wave function of the 4D Higgs boson is given by 
$u_H(y) = [2k/(z_L^2 - 1)]^{1/2}e^{2ky}$ for $0 \leq y \leq L$ and 
$u_H(-y) = u_H(y) = u_H(y + 2L)$. $g_w = g_A/\sqrt{L}$ 
is the dimensionless 4D $SU(2)_L$ coupling.

\section{$H\to Z\gamma$}
The Higgs decay processes  $H\to \gamma\gamma$ and $H\to Z\gamma$ are absent
at the tree level and occur at the one loop level. In the $SO(5)\times U(1)$ GHU
not only the $W$ boson, quarks and leptons in the SM but also their KK modes and 
additional gauge bosons and dark fermions contribute to the processes. 
We show that these corrections are finite and small in the $SO(5)\times U(1)$ GHU,
thanks to the cancellation among them.

\subsection{$\Gamma (H\to\gamma\gamma)$}
We shortly review the decay process $H\to\gamma\gamma$ in the $SO(5)\times U(1)$ 
GHU model.\cite{Funatsu:2013ni} 
We follow the notation of Ref.\ \cite{Gunion:1989we}. The decay rate in the SM is given by 
\begin{align}
	\Gamma ( H \to \gamma \gamma )_\SM = 
	\frac{\alpha^2 g_w^2}{1024 \pi^3}\frac{m_H^3}{m_W^2} \, 
	\Big| \sum_i N_{c\, i}e_i^2 F_i (\tau_i) \Big| ^2 ~,~~~
	\tau_i = \frac{4 m_i^2}{m_H^2} ~, 
	\label{2gamma}
\end{align}
where $N_{c\, i}$ is the number of the color degrees of freedom and $e_i$ is the electromagnetic charge in units of $e$.
Functions $F_1(\tau)$ and $F_{1/2}(\tau)$ are assigned for gauge bosons and fermions, respectively, and defined by
\beqn
&&\hskip -1.cm
F_1(\tau) =2+3\tau  +3\tau(2-\tau) f(\tau) ~,\cr
\noalign{\kern 5pt}
&&\hskip -1.cm
F_{1/2} (\tau) =-2\tau  [1+(1-\tau) f(\tau)] ~,\cr
\noalign{\kern 10pt}
&&\hskip -1.cm
f(\tau)= \begin{cases}
\Big[\sin^{-1}\big(\sqrt{1/\tau}\big) \Big]^2 & \text{for } \tau \geq 1 ~, \cr
\noalign{\kern 5pt}
- \frac{1}{4} \bigg[ \ln \myfrac{1 + \sqrt{1-\tau}}{1 - \sqrt{1-\tau}} - i\pi \bigg]^2
 & \text{for } \tau < 1 ~. \end{cases}
\label{Ffunction}
\eeqn
In the large $\tau$ limit   $F_{1/2}\to-\frac{4}{3}$ and $ F_1\to7$.

In the $SO(5)\times U(1)$ GHU model, 
KK numbers are conserved by the electromagnetic interactions. 
The decay rate in the GHU becomes
\begin{align} 
	\Gamma ( H \to \gamma \gamma ) 
	&= \frac{\alpha^2g_w^2}{1024 \pi^3}\frac{m_H^3}{m_W^2} 
	\left|\mathcal{F}_W+\frac{4}{3}\mathcal{F}_t+n_F\mathcal{F}_F\right|^2, 
\label{2gamma1}
\end{align}
where
\begin{align}
	\mathcal{F}_W=\sum_{n=0}^{\infty}
	\frac{g_{HW^\n W^\n }}{g_wm_W}\frac{m_W^2}{m_{W^\n}^2} F_1(\tau_{W^\n})
	&=\sum_{n=0}^{\infty}
	I_{W^\n}\frac{m_W}{m_{W^\n}} \cos\theta_HF_1(\tau_{W^\n}),\nonumber\\
	\mathcal{F}_t=\;\;\sum_{n=0}^{\infty}
	\frac{y_{Ht^\n t^\n }}{y_t^\text{SM}}\frac{m_t}{m_{t^\n}} F_{1/2}(\tau_{t^\n})\;\;
	&=\sum_{n=0}^{\infty}
	I_{t^\n }\frac{m_t}{m_{t^\n}} \cos\theta_HF_{1/2}(\tau_{t^\n}),\nonumber\\
	\mathcal{F}_F=\sum_{n=1}^{\infty}
	\frac{y_{HF^\n F^\n }}{y_t^\text{SM}}\frac{m_t}{m_{F^\n}} F_{1/2}(\tau_{F^\n})
	&=\sum_{n=1}^{\infty}
	I_{F^\n }\frac{m_t}{m_{F^\n}} \sin\frac{\theta_H}{2}F_{1/2}(\tau_{F^\n}).
\label{2gamma2}
\end{align}
Here $I_{W^\n}$, $I_{t^\n}$ and $I_{F^\n}$ are defined as
$I_{W^{(n)}}=g_{HW^\n W^\n}/g_wm_{W^{(0)}}\cos\theta_H$, 
$I_{t^\n}=y_{Ht^\n t^\n }/y_t^\text{SM}\cos\theta_H$ and $I_{F^{(n)}}=y_{H F^\n F^\n }/y_t^\text{SM}\sin\frac{\theta_H}{2}$.
Contributions from light quarks and leptons and their KK modes are negligible.

\begin{figure}
\centering
\includegraphics[bb=0 0 415 214,width=12cm]{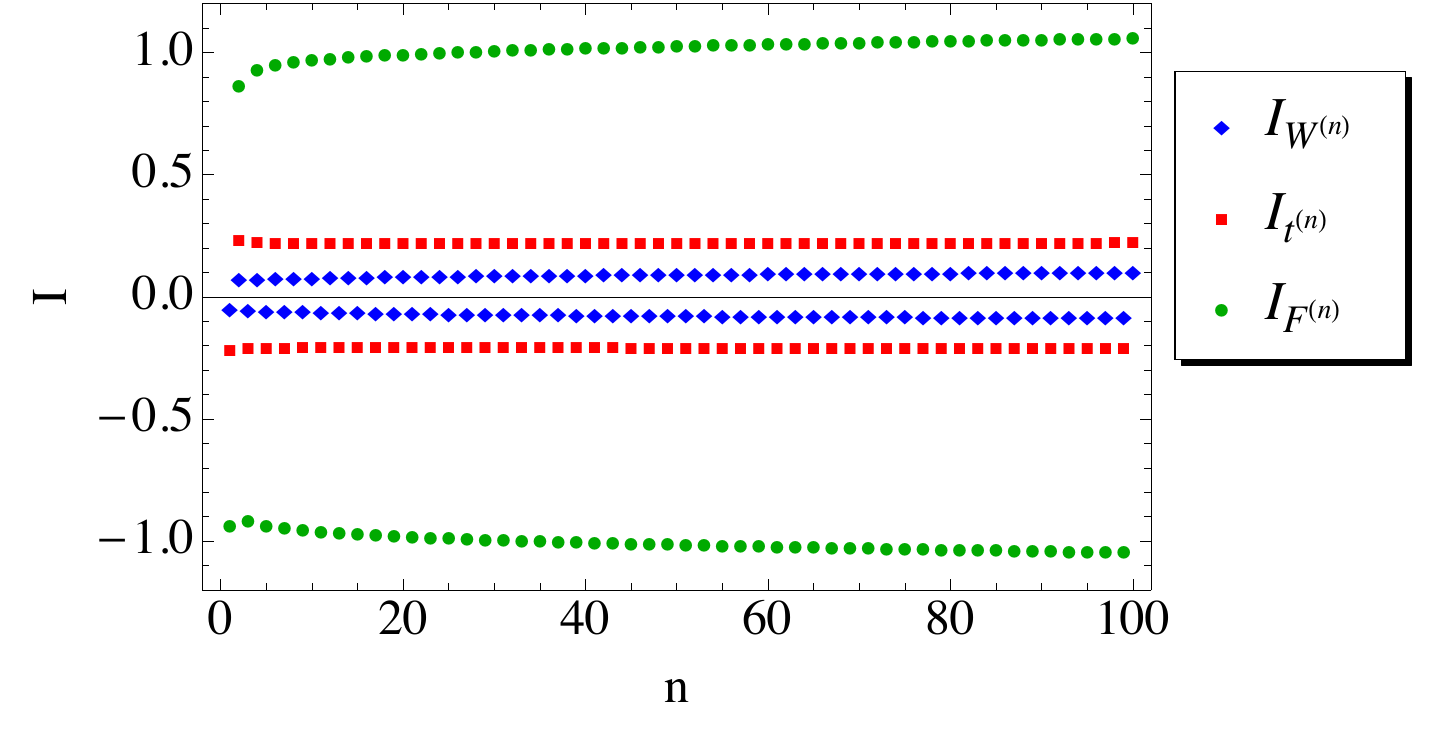}
\caption{Behaviors of $I_{W^{(n)}}=g_{HW^\n W^\n}/g_wm_{W^\n}\cos\theta_H$, 
$I_{t^{(n)}}=y_{Ht^\n t^\n }/y_t^\text{SM}\cos\theta_H$ and 
$I_{F^{(n)}}=y_{H F^\n F^\n }/y_t^\text{SM}\sin\frac{\theta_H}{2}$
 in the case of $n_F=4$, $z_L=10^5$ for which $\theta_H = 0.1153$. }
\label{IWtF}
\end{figure}

In Fig.~\ref{IWtF}, the values of $I_{W^\n}$, $I_{t^\n}$ and $I_{F^\n}$ by the numerical calculation 
in the $N_F=4$ and $\theta_H=0.1153$ case are shown.
They approximately  behave as 
\begin{align}
I_{W^{(n)}}\simeq (-1)^n \big\{0.0759 - 0.0065 \ln n + 0.0022 (\ln n)^2 \big\} ~, \nonumber\\
I_{t^{(n)}}\simeq (-1)^n \big\{0.2304 - 0.0108 \ln n + 0.0017 (\ln n)^2 \big\} ~, \nonumber\\
I_{F^{(n)}}\simeq (-1)^n \big\{1.0341 - 0.0457 \ln n + 0.0108 (\ln n)^2 \big\} ~~
\end{align}
for $50 \le n \le 200$. 
Note that the sign alternates in $n$. 
As the masses of the KK modes of the $W$ boson,  top quark and  dark fermion
are $m_n\simeq n\cdot m_\KK/2$,  the sum in each $\mathcal{F}$ behaves as
$\sum (-1)^n (\ln n)^\alpha /n$ ($\alpha =0,1,2$) and converges.  
Moreover the contributions from $n \ge 1$ are suppressed by the ratio of
the electroweak scale to the KK scale.
Hence the ratio of $\mathcal{F}$ to the  zero-mode contribution becomes
\begin{align}
	\frac{\mathcal{F}_W}{\mathcal{F}_{W^{(0)} \text{only}}}=& \quad 0.9997 ~, \cr
	\frac{\mathcal{F}_t}{\mathcal{F}_{t^{(0)} \text{only}}}=& \quad 0.9983 ~, \cr
	\frac{\mathcal{F}_F}{\mathcal{F}_{t^{(0)} \text{only}}}=& -0.0032 ~,  
\label{2gamma3}
\end{align}
for $\theta_H = 0.114$ and $n_F=4$. 
The ratio of the amplitude to that with only zero modes is 
\begin{align}
	\frac{\mathcal{F}_W+\frac{4}{3}\mathcal{F}_t+4\mathcal{F}_F}
	{\mathcal{F}_{W^{(0)} \text{only}}+\frac{4}{3}\mathcal{F}_{t^{(0)} \text{only}}}=1.0027.
\end{align}
It is found that  the contributions of the KK modes are less than 1\% and negligible. 
The couplings of the zero modes are approximately given by 
$g_{HW W} \simeq g_{HW W}^{\SM} \cos \theta_H = g_w m_W \cos \theta_H $ and
$y_t \simeq y_t^{\SM} \cos \theta_H$. 
Therefore the decay rate in the GHU is approximately $\cos^2\theta_H$ times that in the SM.
The deviation  from the SM amounts only  $1\%$ for $\theta_H \sim 0.1$.

\subsection{Gauge boson loops}
The decay process $H\to Z\gamma$ is more involved than $H\to \gamma \gamma$.
The KK number need not be conserved at the $Z$ and $H$ vertices, and $W_R^{(n)}$
also participates.
\begin{figure}[thb]
	\begin{minipage}{0.5 \hsize}\centering
		\includegraphics[bb=0 0 285 132, width=6.5cm]{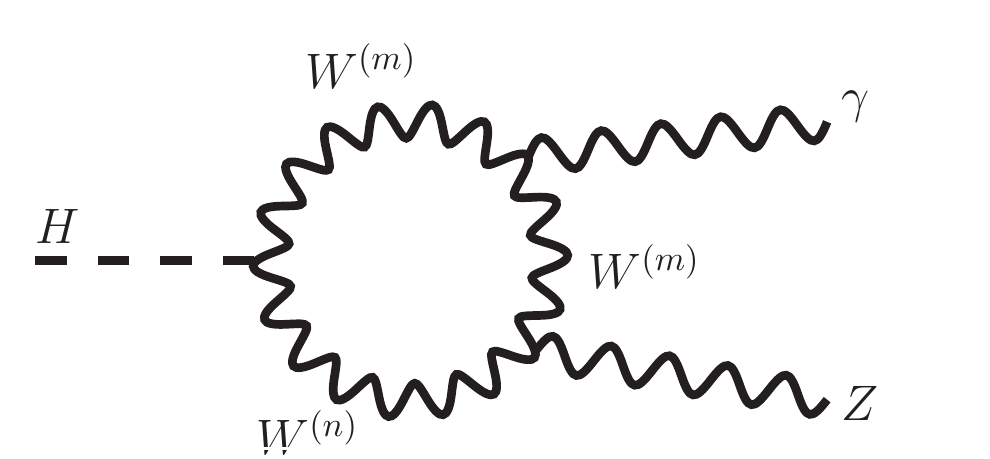}\subcaption{}
	\end{minipage}
	\begin{minipage}{0.5 \hsize}\centering
		\includegraphics[bb=0 0 285 132, width=6.5cm]{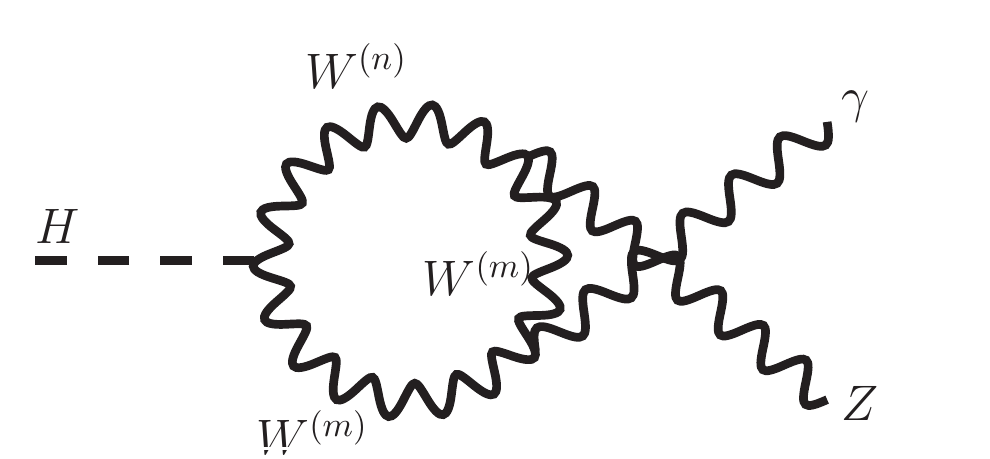}\subcaption{}
	\end{minipage}
	\begin{minipage}{0.5 \hsize}\centering
		\includegraphics[bb=0 0 285 132, width=6.5cm]{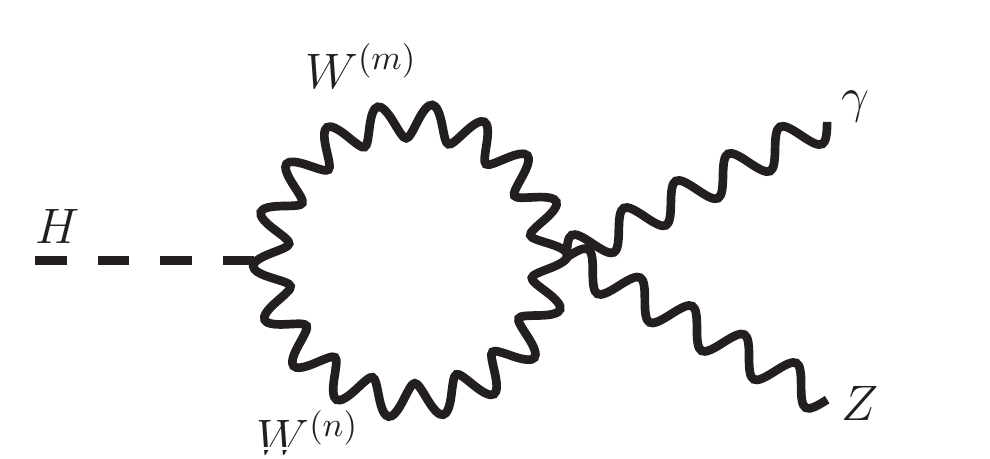}\subcaption{}
	\end{minipage}
	\begin{minipage}{0.5 \hsize}\centering
		\includegraphics[bb=0 0 285 136, width=6.5cm]{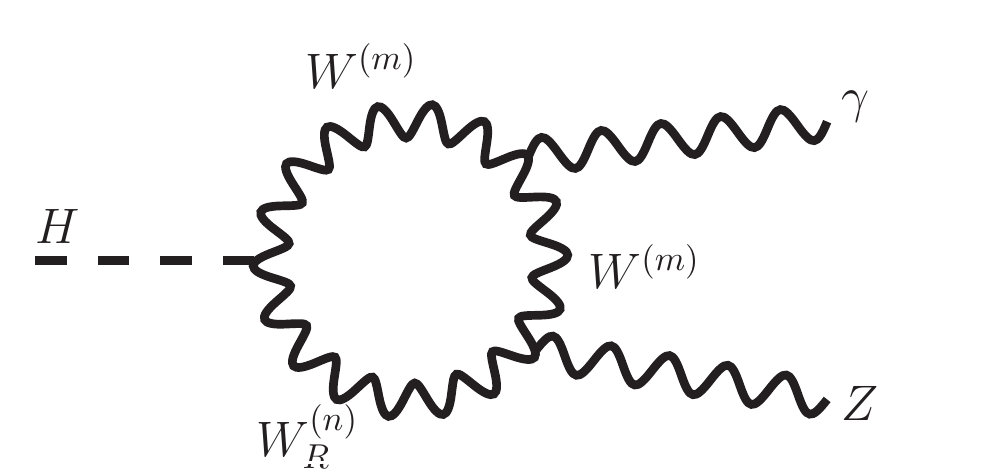}\subcaption{}
	\end{minipage}
	\begin{minipage}{0.5 \hsize}\centering
		\includegraphics[bb=0 0 285 136, width=6.5cm]{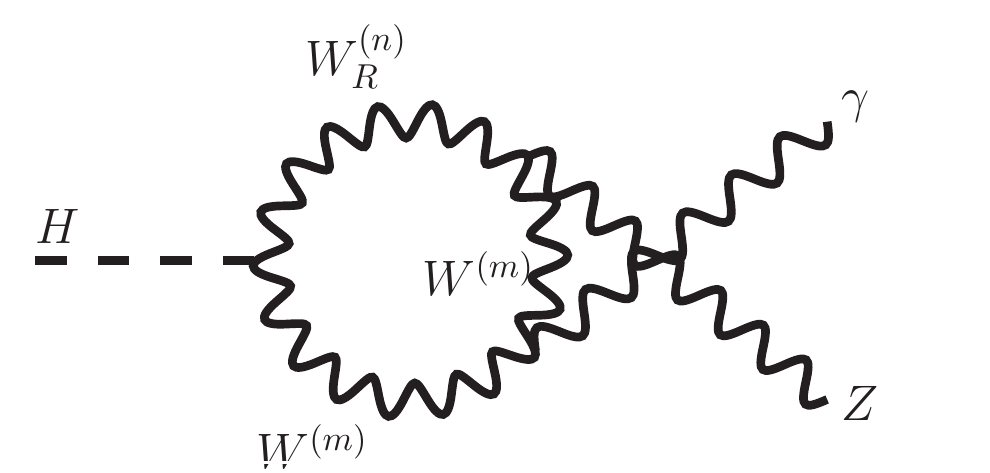}\subcaption{}
	\end{minipage}
	\begin{minipage}{0.5 \hsize}\centering
		\includegraphics[bb=0 0 285 136, width=6.5cm]{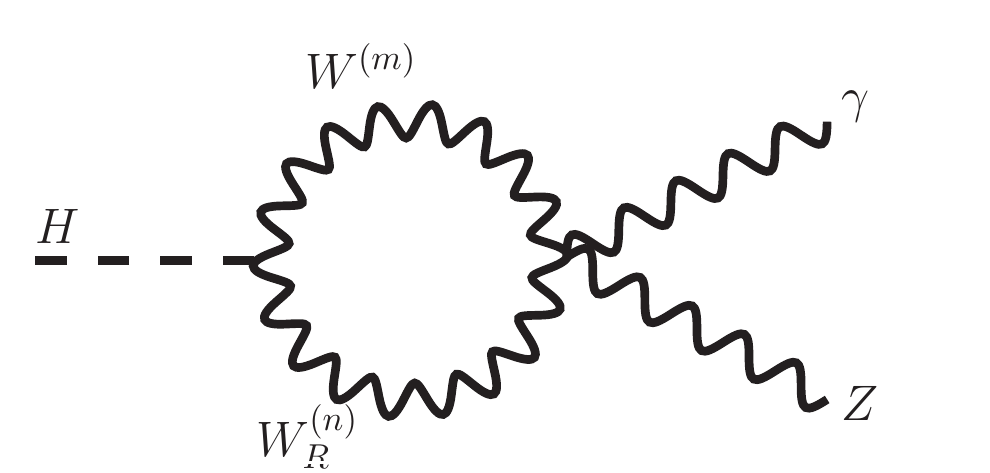}\subcaption{}
	\end{minipage}  
	\caption{
	The gauge boson loop processes for $H\to Z\gamma$ in the $SO(5) \times U(1)$ 
	gauge-Higgs unification. $W_R$ is the $SU(2)_R$ gauge boson and has no zero mode. 
	Since $HW_R^\m W_R^\n$ couplings vanish, there are no diagrams involving two or more $W_R$' s.}
	\label{gauge boson diagrams}
\end{figure}
The gauge boson loop processes for $H\to Z\gamma$  are shown in Fig.~\ref{gauge boson diagrams}. 
We note that $HW_R^\m W_R^\n$ couplings vanish. 
The amplitude of $W$ boson loop Figs.~\ref{gauge boson diagrams}(a)(b)(c) is given in the unitary gauge  by
\beqn
&&\hskip -1.cm
i\mathcal{M}^{(a)}_{W^\m,W^\n}+i\mathcal{M}^{(b)}_{W^\m,W^\n}+i\mathcal{M}^{(c)}_{W^\m,W^\n}\cr
\noalign{\kern 10pt}
&&\hskip -1.cm
= eg_{HW^\m W^\n}g_{ZW^\m W^\n}\epsilon^*_\mu(k_1)\epsilon^*_\nu(k_2)  
 \int\frac{d^4p}{(2\pi)^4} \, D_{\tau \alpha} (p, m_{W^\m}) 
{D_{\sigma}}^\tau (p-k_1 -k_2, m_{W^\n})  \cr
\noalign{\kern 5pt}
&&\hskip -.0cm
\times  \Big[ 2 D_{\beta \rho} (p-k_1 , m_{W^\m})
\big\{ 2\eta^{\alpha\beta}p^\mu-\eta^{\beta\mu}(p-2k_1)^\alpha-\eta^{\alpha\mu}(p+k_1)^\beta \big\} \cr
\noalign{\kern 5pt}
&&\hskip 1.5cm
\times \big\{ 2\eta^{\rho\sigma}(p-k_1)^\nu-\eta^{\sigma\nu}(p-k_1-2k_2)^\rho
     -\eta^{\rho\nu}(p-k_1+k_2)^\sigma  \big\} \cr
\noalign{\kern 5pt}
&&\hskip 0.5cm
- (2\eta^{\mu\nu} \eta^{\alpha\sigma} - \eta^{\mu\alpha} \eta^{\nu\sigma}
-\eta^{\mu\sigma} \eta^{\nu\alpha}) \Big] ~, \cr
\noalign{\kern 10pt}
&&\hskip 0.cm
D_{\mu \nu} (p, m) = \Big( \eta_{\mu\nu} - \frac{p_\mu p_\nu}{m^2} \Big) \frac{1}{p^2 - m^2 + i \ep} ~,
\label{gauge boson amplitude}
\eeqn
where $k_1$ and $k_2$ are the photon and the $Z$ boson momenta, respectively.
The amplitude \eqref{gauge boson amplitude} itself is divergent. 
However, by adding the $m \leftrightarrow n$ diagrams and using 
$g_{HW^\m W^\n}=g_{HW^\n W^\m}$ and  $g_{ZW^\m W^\n}=g_{ZW^\n W^\m}$, 
the amplitude becomes 
\begin{align}
 i & \big\{ \mathcal{M}^{(a)}_{W^\m,W^\n}+ \mathcal{M}^{(b)}_{W^\m,W^\n} + \mathcal{M}^{(c)}_{W^\m,W^\n}
 +(m \longleftrightarrow n)\big\} \nonumber\\
   =&eg_{HW^\m W^\n}g_{ZW^\m W^\n}\epsilon^*_\mu(k_1)\epsilon^*_\nu(k_2)
	\left(\eta^{\mu\nu}-\frac{k_2^\mu k_1^\nu}{k_1\cdot k_2} \right)\frac{i}{16\pi^2}\frac{1}{m_{W^\m}^2 m_{W^\n}^2}
	\nonumber\\
    &\times\Big\{\Big(m_{W^\m}^4+m_{W^\n}^4+10 m_{W^\m}^2m_{W^\n}^2\Big)
    E_+(m_{W^\m},m_{W^\n})
	\nonumber\\
	&\quad+\Big((m_{W^\m}^2+m_{W^\n}^2)(m_H^2-m_Z^2)-m_H^2m_Z^2\Big)E_- (m_{W^\m},m_{W^\n})
	\nonumber\\&\quad
	-\left(4m_{W^\m}^2 m_{W^\n}^2 (m_H^2-m_Z^2)+2m_Z^4 (m_{W^\m}^2+m_{W^\n}^2)\right) \nonumber\\
	&\qquad\times \left(C_0(m_{W^\m},m_{W^\n})+C_0(m_{W^\n},m_{W^\m})\right)\Big\}\label{WdiagResult}
\end{align}
where 
\begin{align}
C_0(m_1^2,m_2^2) &\equiv C_0(0,m_H^2,m_Z^2,m_1^2,m_1^2,m_2^2) ~,\nonumber\\
E_\pm (m_1,m_2)&\equiv 1+\frac{m_Z^2}{m_H^2-m_Z^2}
\left\{B_0(m_H^2,m_1^2,m_2^2)-B_0(m_Z^2,m_1^2,m_2^2)\right\} \nonumber\\
&\quad\pm \left\{m_1^2 C_0(m_1^2,m_2^2)+m_2^2 C_0(m_2^2,m_1^2)\right\}~,\end{align}
with the Passarino-Veltman functions\cite{Passarino:1978jh, Denner:1991kt} defined by
\beqn
&&\hskip -1.cm
B_0(k^2,m_1^2,m_2^2)
	\equiv\frac{(2\pi)^{4-D}}{i\pi^2}\int d^Dq \frac{1}{(q^2-m_1^2)\{(q+k)^2-m_2^2\}} ~,  \cr
\noalign{\kern 10pt}
&&\hskip -1.cm
C_0(k_1^2,(k_1 - k_2)^2,k_2^2,m_1^2,m_2^2,m_3^2) \cr
\noalign{\kern 10pt}
&&\hskip 0.cm
\equiv\frac{1}{i\pi^2}\int d^Dq \frac{1}{(q^2-m_1^2)\{(q+k_1)^2-m_2^2\}\{(q+k_2)^2-m_3^2\}} ~,
\label{PV1}
\eeqn
For all $C_0$ and $E_\pm$, the $D\rightarrow 4$ limit has been safely taken, so that the amplitude \eqref{WdiagResult} is finite.

\begin{table}[hbtp]\centering \footnotesize
\caption{$J_{W^\m W^\n}$ defined in \eqref{JWW1} 
is shown for $0\le m,n\le 7$ and for $101\le m,n\le 108$ in the $N_F=4$, $z_L=10^5$ case. 
Only the values larger than that of $O(10^{-4})$ are shown with three significant figures.} 
\label{tableJWW-1}
\begin{tabular}{|c|cccccccc|}\hline
	& 0 & 1 & 2 & 3 & 4 & 5 & 6 & 7  \\ \hline
  0 &     1.00     & $O(10^{-4})$ & $O(10^{-9})$ & $O(10^{-6})$ & $O(10^{-11})$ & $O(10^{-8})$ 
    & $O(10^{-12})$& $O(10^{-9})$ \\
  1 & $O(10^{-4})$ &   -0.0580    &  \ 0.0595    & $O(10^{-6})$ & $O(10^{-5})$ & $O(10^{-10})$
    & $O(10^{-7})$ & $O(10^{-9})$ \\
  2 & $O(10^{-9})$ &    0.0595    &  \ 0.0218    &    -0.0413   & $O(10^{-8})$ & $O(10^{-5})$ 
    & $O(10^{-9})$ & $O(10^{-5})$ \\
  3 & $O(10^{-6})$ & $O(10^{-6})$ &   -0.0413    &    -0.0625   &  \ 0.0637    & $O(10^{-6})$
    & $O(10^{-5})$ & $O(10^{-10})$\\
  4 & $O(10^{-11})$& $O(10^{-5})$ & $O(10^{-8})$ &   \ 0.0637   &  \ 0.0226    &   -0.0432
    & $O(10^{-7})$ & $O(10^{-5})$ \\
  5 & $O(10^{-8})$ & $O(10^{-10})$& $O(10^{-5})$ & $O(10^{-6})$ &   -0.0432    &   -0.0652
    &  \ 0.0648    & $O(10^{-6})$ \\
  6 & $O(10^{-12})$& $O(10^{-7})$ & $O(10^{-9})$ & $O(10^{-5})$ & $O(10^{-7})$ &  \ 0.0648
    &  \ 0.0233    &   -0.0434    \\
  7 & $O(10^{-9})$ & $O(10^{-9})$ & $O(10^{-5})$ & $O(10^{-10})$& $O(10^{-5})$ & $O(10^{-6})$
    &   -0.0434    &   -0.0673    \\
\hline\end{tabular}
\vskip 8pt
\begin{tabular}{|c|cccccccc|}\hline
    & 101 & 102 & 103 & 104 & 105 & 106 & 107 & 108 \\ \hline
101 &   -0.0932    &  \ 0.0705    & $O(10^{-6})$ & $O(10^{-4})$ & $O(10^{-12})$ 
    & $O(10^{-5})$ & $O(10^{-8})$ & $O(10^{-6})$  \\
102 &  \ 0.0705    &  \ 0.0328    &   -0.0406    & $O(10^{-6})$ & $O(10^{-4})$ 
    & $O(10^{-12})$& $O(10^{-5})$ & $O(10^{-9})$  \\
103 & $O(10^{-6})$ &   -0.0406    &   -0.0934    &  \ 0.0706    & $O(10^{-6})$ 
    & $O(10^{-4})$ & $O(10^{-12})$& $O(10^{-5})$  \\
104 & $O(10^{-4})$ & $O(10^{-6})$ &  \ 0.0706    &  \ 0.0329    &   -0.0405
    & $O(10^{-6})$ & $O(10^{-4})$ & $O(10^{-12})$ \\
105 & $O(10^{-12})$& $O(10^{-4})$ & $O(10^{-6})$ &   -0.0405    &   -0.0937 
    &  \ 0.0706    & $O(10^{-6})$ & $O(10^{-4})$  \\
106 & $O(10^{-5})$ & $O(10^{-12})$& $O(10^{-4})$ & $O(10^{-6})$ &  \ 0.0706
    &  \ 0.0330    &   -0.0405    & $O(10^{-6})$  \\
107 & $O(10^{-8})$ & $O(10^{-5})$ & $O(10^{-12})$& $O(10^{-4})$ & $O(10^{-6})$
    &   -0.0405    &   -0.0940    &  \ 0.0707     \\
108 & $O(10^{-6})$ & $O(10^{-9})$ & $O(10^{-5})$ & $O(10^{-12})$& $O(10^{-4})$ 
    & $O(10^{-6})$ &  \ 0.0707    &  \ 0.0331     \\
\hline
\end{tabular}\end{table}

To obtain the amplitude quantitatively,  the numerical values of the couplings $g_{HW^\m W^\n}$ 
and $g_{ZW^\m W^\n}$ have to be evaluated.
The details of evaluation and results are summarised in Appendix.
It is convenient to define
the dimensionless coupling $J_{W^\m W^\n}$ by
\beeq
J_{W^\m W^\n} \equiv \frac{g_{HW^\m W^\n} g_{ZW^\m W^\n}}
	{g_w^2\cos\theta_W\cos\theta_H \sqrt{m_{W^\m}m_{W^\n}}} ~.
\label{JWW1}
\eneq
The value of $J_{W^\m W^\n}$ is tabulated in Table~\ref{tableJWW-1}.  
It is  seen  that $J_{W^\m W^\n}$ with $|m-n|\ge 2$ is smaller than $J_{W^\n W^\n}$
by a factor $10^{-2}$, 
whereas  $J_{W^{(n\pm 1)} W^\n}$ and  $J_{W^\n W^\n}$ are of the same order.
\begin{figure}
	\centering
	\includegraphics[bb=0 0 415 187, width=12cm]{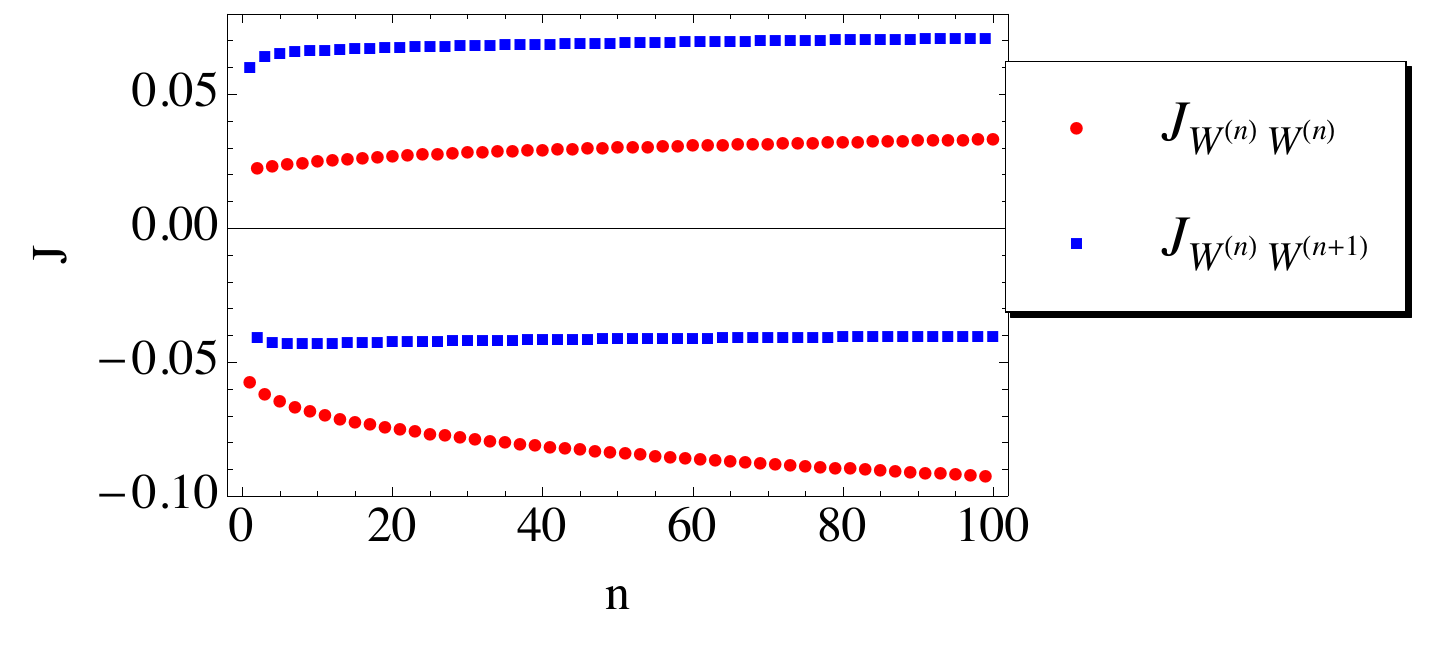}
	\caption{$J_{W^\n W^\n}$ and $J_{W^\n W^{(n+1)}}$ are plotted for $1\le n \le 100$ 
	in the $N_F=4$, $z_L=10^5$ case. The red circles and blue squares represent   
	$J_{W^\n W^\n}$ and $J_{W^\n W^{(n+1)}}$, respectively.}\label{JWW-plot}
\end{figure}
$J_{W^\n W^\n}$ and $J_{W^\n W^{(n+1)}}$ are plotted in Fig.~\ref{JWW-plot} for $1\le n \le 100$ in the $N_F=4$, $z_L=10^5$ case. 
$J_{W^\n W^\n}$ and $J_{W^\n W^{(n+1)}}$ for $101 \le n \le 200$ are 
approximately given by 
\beqn
&&\hskip -1.cm
J_{W^\n W^\n}   \simeq -0.0272+0.00320 (\ln n)-0.00083 (\ln n)^2 \cr
\noalign{\kern 5pt}
&&\hskip 1.cm
+(-1)^{n-1}\left( -0.0563+0.00654 (\ln n)-0.00173 (\ln n)^2\right) ~,  \cr 
\noalign{\kern 10pt}
&&\hskip -1.cm
J_{W^\n W^{(n+1)}}   \simeq 0.0135-0.00160 (\ln n)+0.00041 (\ln n)^2  \cr
\noalign{\kern 5pt}
&&\hskip 1.cm
+(-1)^{n-1}    \left( 0.0567-0.00106 (\ln n)+0.00018 (\ln n)^2\right) ~.
\label{JWW2}
\eeqn

Next we examine the asymptotic behavior of the amplitude for $m,n\gg 1$.
The whole amplitude of the $W$ boson loop is
\begin{align}
i\mathcal{M}_W=
&\frac{i}{2}\sum_{m, n}^\infty\Big\{ \mathcal{M}^{(a)}_{W^\m,W^\n}
+\mathcal{M}^{(b)}_{W^\m,W^\n} + \mathcal{M}^{(c)}_{W^\m,W^\n} + ( n \longleftrightarrow m ) \Big\}  ~.
\label{ampW2}
\end{align}
The diagonal part of the amplitude in \eqref{ampW2} for $n\gg 1$ is rewritten as 
\begin{align}
	&i\mathcal{M}^{(a)}_{W^\n,W^\n}+i\mathcal{M}^{(b)}_{W^\n,W^\n}+i\mathcal{M}^{(c)}_{W^\n,W^\n}\nonumber\\
	=&eg_w^2\cos\theta_W\cos\theta_H\epsilon^*_\mu(k_1)\epsilon^*_\nu(k_2)
	\left(\eta^{\mu\nu}-\frac{k_2^\mu k_1^\nu}{k_1\cdot k_2} \right)\frac{i}{16\pi^2}
    \frac{J_{W^\n W^\n}}{2m_{W^\n}^3}\nonumber\\
	&\times\bigg\{-\frac{m_H^2-m_Z^2}{2m_{W^\n}^2}
	\Big(12 m_{W^\n}^4+2m_{W^\n}^2(m_H^2-m_Z^2)-m_H^2m_Z^2\Big)
	I_1(\tau_{W^\n},\lambda_{W^\n})\nonumber\\
	&\qquad+4\Big(4m_{W^\n}^2(m_H^2-m_Z^2)-m_H^2m_Z^2+ m_Z^4\Big)I_2(\tau_{W^\n},\lambda_{W^\n})\bigg\},
\end{align}
where 
\begin{align}
	&I_1(a,b)=\frac{ab}{2(a-b)}+\frac{a^2b^2}{2(a-b)^2}\big[f(a)-f(b)\big]  
	+\frac{a^2b}{(a-b)^2}\big[g(a)-g(b) \big] ,\cr
	&I_2(a,b)=-\frac{ab}{2(a-b)}[f(a)-f(b)]~ , \cr
	&g(\tau)= \begin{cases}
	\sqrt{\tau-1}\sin^{-1}\big(\sqrt{1/\tau}\big) & \text{for } \tau \geq 1 ~, \cr
	\noalign{\kern 5pt}
	\frac{1}{2}\sqrt{1-\tau} \bigg[ \ln \myfrac{1 + \sqrt{1-\tau}}{1 - \sqrt{1-\tau}} - i\pi \bigg]
	& \text{for } \tau < 1 ~. \end{cases}
\label{Ifunction}
\end{align}
and $\lambda_{i}\equiv 4m_{i}^2/m_Z^2$. $\tau_i$ and $f(a)$ are defined in \eqref{2gamma} and \eqref{Ffunction}.
Here, we have used 
\begin{align}
&\frac{m_Z^2}{m_H^2-m_Z^2}\left(B_0(m_H^2,m_{W^\n}^2,m_{W^\n}^2)
-B_0(m_Z^2,m_{W^\n}^2,m_{W^\n}^2)\right) \cr
&\qquad =-1-\frac{m_H^2-m_Z^2}{2m_{W^\n}^2}I_1(\tau_{W^\n},\lambda_{W^\n})+2I_2(\tau_{W^\n},\lambda_{W^\n}) ~, \cr
&C_0(0,m_H^2,m_Z^2,m_{W^\n}^2,m_{W^\n}^2,m_{W^\n}^2)
=-\frac{1}{m_{W^\n}^2}I_2(\tau_{W^\n},\lambda_{W^\n}) ~.
\label{BCfunction}
\end{align}
The functions $I_1, I_2$ in \eqref{BCfunction} approach constants for $m_{W^{(n)}}\to \infty$.
Since $J_{W^\m W^\n}$ for $|m-n|\ge 2$ are negligible comparing to $J_{W^\m W^\n}$ for $|m-n|\le 1$,  
only the amplitude for $|m-n|\le 1$ need to be retained. 
For $n \gg 1$ $m_{W^{(n\pm 1)}}\simeq m_{W^\n}$ so that
\beqn
&&\hskip -1.cm
\frac{1}{2}\sum_{m}\Big(
i\mathcal{M}^{(a)}_{W^\m,W^\n}+i\mathcal{M}^{(b)}_{W^\m,W^\n}+i\mathcal{M}^{(c)}_{W^\m,W^\n}
+ ( n \longleftrightarrow m ) \Big) \cr
\noalign{\kern 5pt}
&&\hskip -1.cm
\simeq \; eg_w^2\cos\theta_W\cos\theta_H\epsilon^*_\mu(k_1)\epsilon^*_\nu(k_2)
\left(\eta^{\mu\nu}-\frac{k_2^\mu k_1^\nu}{k_1\cdot k_2} \right)\frac{i}{16\pi^2}\cr
\noalign{\kern 5pt}
&&\hskip -0.cm
\times\frac{1}{2m_{W^\n}^3}\left(J_{W^\n W^\n}+J_{W^{(n+1)} W^\n}+J_{W^{(n-1)} W^\n}\right)\cr
\noalign{\kern 5pt}
&&\hskip -0.cm
\times \bigg\{-\frac{m_H^2-m_Z^2}{2m_{W^\n}^2}\Big(12 m_{W^\n}^4+2m_{W^\n}^2(m_H^2-m_Z^2)
-m_H^2m_Z^2\Big) I_1(\tau_{W^\n},\lambda_{W^\n})\cr
\noalign{\kern 5pt}
&&\hskip -0.cm
\qquad+4\Big(4m_{W^\n}^2(m_H^2-m_Z^2)-m_H^2m_Z^2+ m_Z^4\Big)
I_2(\tau_{W^\n},\lambda_{W^\n})\bigg\}\cr
\noalign{\kern 5pt}
&&\hskip -1.cm
\approx  \;\text{const.}\times \frac{1}{n}\left(J_{W^\n W^\n}+J_{W^{(n+1)} W^\n}+J_{W^{(n-1)} W^\n}\right) .
\label{ampW3}
\eeqn
Therefore the large $n$ part of the sum in the whole amplitude of $W$ boson loop 
is approximated as
\begin{align}
	i\mathcal{M}_W \approx \sum_n^\infty\text{const.}\times 
	\frac{1}{n}\left(J_{W^\n W^\n}+J_{W^{(n+1)} W^\n}+J_{W^{(n-1)} W^\n}\right) .
\end{align}
Even though the sums $\sum J_{W^\n W^\n}/n $ and $\sum J_{W^{(n\pm1)} W^\n}/n $ 
diverge individually, the combination 
$\sum\left(J_{W^\n W^\n}+J_{W^{(n+1)} W^\n}+J_{W^{(n-1)} W^\n}\right)/n$  converges
since
\begin{align}
	&J_{W^\n W^\n}+J_{W^{(n+1)} W^\n}+J_{W^{(n-1)} W^\n}\nonumber\\
	&\simeq (-1)^{n-1}\big\{ -0.0563+0.00654 (\ln n)-0.00173 (\ln n)^2\big\} ~.
\end{align}

Next we consider the amplitudes which contain $W_R$ in the loop.
The amplitude $i\mathcal{M}^{(d)}_{W^\m,W_R^\n}+i\mathcal{M}^{(e)}_{W^\m,W_R^\n}+i\mathcal{M}^{(f)}_{W^\m,W_R^\n}$ 
is obtained from (\ref{gauge boson amplitude}) by  replacing $W^\n$ by $W_R^\n $.
The dimensionless coupling $J_{W^\m W^\n_R}$ is also defined as 
\begin{align}
	J_{W^\m W^\n_R} \equiv \frac{g_{HW^\m W^\n_R} g_{ZW^\m W^\n_R}}
	{g_w^2\cos\theta_W \sqrt{m_{W^\m}m_{W^\n_R}}} ~.
\label{JWWR1}
\end{align}

\begin{figure}
	\centering
	\includegraphics[bb=0 0 550 247, width=13cm]{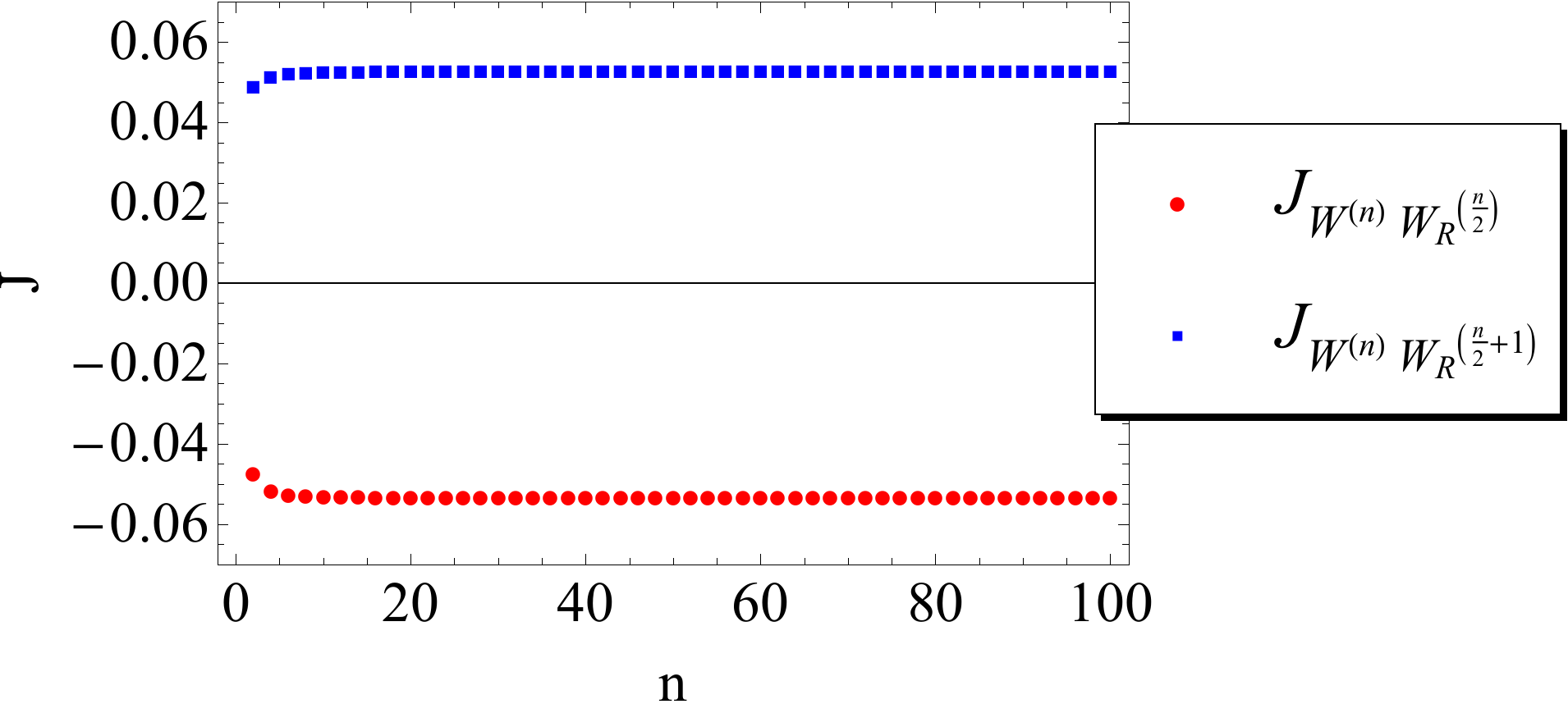}
\caption{$J_{W^\n W_R^{(n/2)}}$ and $J_{W^\n W_R^{(n/2+1)}}$ 
are plotted for $1\le n \le 100$ in the $N_F=4$, $z_L=10^5$ case. 
The red circles and  blue squares represent 
$J_{W^\n W_R^{(n/2)}}$ and $J_{W^\n W_R^{(n/2+1)}}$, respectively.} 
\label{JWWR-plot}
\end{figure}

\begin{table}[tbhp]\centering \footnotesize
\caption{$J_{W^\m W^\n_R}$ in \eqref{JWWR1} 
is shown for $0\le m \le 7$, $1\le n \le 4$ and for $101\le m\le 108$, $51\le n \le 55$
 in the $N_F=4$, $z_L=10^5$ case. 
Only values larger than $O(10^{-4})$ are shown explicitly with three significant figures.}
\label{tableJWWR}
\begin{tabular}{|cc|cccccccc|}\hline
	&&&&&$m$&&&&\\
	&& 0 & 1 & 2 & 3 & 4 & 5 & 6 & 7  \\ \hline
    & 1 & $O(10^{-4})$ & $O(10^{-4})$ &   -0.0470    & $O(10^{-7})$ & $O(10^{-5})$ 
        & $O(10^{-10})$ & $O(10^{-6})$ & $O(10^{-10})$ \\
    & 2 & $O(10^{-6})$ & $O(10^{-6})$ &  \ 0.0491    & $O(10^{-4})$ &   -0.0514     
        & $O(10^{-7})$ & $O(10^{-4})$ & $O(10^{-10})$ \\
$n$ & 3 & $O(10^{-8})$ & $O(10^{-9})$ & $O(10^{-4})$ & $O(10^{-6})$ &  \ 0.0517
        & $O(10^{-4})$ &   -0.0523    & $O(10^{-7})$ \\
    & 4 & $O(10^{-8})$ & $O(10^{-9})$ & $O(10^{-5})$ & $O(10^{-9})$ & $O(10^{-5})$ 
        & $O(10^{-6})$ &  \ 0.0524    & $O(10^{-4})$ \\
\hline
\end{tabular}
\vskip 5pt
\begin{tabular}{|cc|cccccccc|}\hline
	&&&&&$m$&&&&\\
	& & 101 & 102 & 103 & 104 & 105 & 106 & 107 & 108 \\ \hline
    & 51 & $O(10^{-4})$ &   -0.0530    & $O(10^{-6})$ & $O(10^{-4})$ & $O(10^{-13})$ 
         & $O(10^{-5})$ & $O(10^{-9})$ & $O(10^{-6})$ \\
    & 52 & $O(10^{-6})$ &  \ 0.0532    & $O(10^{-4})$ &   -0.0530    & $O(10^{-6})$ 
         & $O(10^{-4})$ & $O(10^{-13})$& $O(10^{-5})$ \\
$n$ & 53 & $O(10^{-11})$& $O(10^{-4})$ & $O(10^{-6})$ &  \ 0.0532    & $O(10^{-4})$
         &   -0.0530    & $O(10^{-6})$ & $O(10^{-4})$ \\
    & 54 & $O(10^{-8})$ & $O(10^{-5})$ & $O(10^{-11})$& $O(10^{-4})$ & $O(10^{-6})$
         &  \ 0.0532    & $O(10^{-4})$ &   -0.0530    \\
    & 55 & $O(10^{-12})$& $O(10^{-6})$ & $O(10^{-8})$ & $O(10^{-5})$ & $O(10^{-11})$ 
         & $O(10^{-4})$ & $O(10^{-6})$ &  \ 0.0532    \\
\hline\end{tabular}
\end{table}
As is seen in the Table~\ref{tableJWWR} ,
$J_{W^\m W^\n_R}$ is appreciable only when $m/2-n = 0,1$.
Note that $m_{W_R^\n}\simeq n\cdot m_\KK$ while $m_{W^\n}\simeq n\cdot m_\KK/2$. 
For $m/2-n = 0,1$, $m_{W^\m}\simeq m_{W^\n}\simeq n\cdot m_\KK$ is satisfied.
Hence the whole amplitude of the $W$-$W_R$ boson loop is
\begin{align}
	i\mathcal{M}_{W_R}
	&=\sum_{m, n}^\infty	
	\Big(i\mathcal{M}^{(d)}_{W^\m,W^\n_R}+i\mathcal{M}^{(e)}_{W^\m,W^\n_R}+i\mathcal{M}^{(f)}_{W^\m,W^\n_R}\nonumber\\
	&\quad\qquad+i\mathcal{M}^{(d)}_{W^\n_R,W^\m}+i\mathcal{M}^{(e)}_{W^\n_R,W^\m}+i\mathcal{M}^{(f)}_{W^\n_R,W^\m}\Big)
	\nonumber\\
	&\approx\sum_n^\infty\text{const.}\times \frac{1}{n}
	\left(J_{W^\n W^{(n/2)}_R}+J_{W^{(n)} W^{(n/2+1)}_R}\right) ~.
\end{align}
$J_{W^\n W_R^{(n/2)}}$ and $J_{W^\n W_R^{(n/2+1)}}$ are plotted 
in Fig.~\ref{JWWR-plot} for $1\le n \le 100$ in the $N_F=4$, $z_L=10^5$ case. 
$J_{W^\n W_R^{(n/2)}}$ and $J_{W^\n W_R^{(n/2+1)}}$
are  approximated in this range by
\begin{align}
	J_{W^\n W_R^{(n/2)}}   &\simeq - 0.0530 -0.000018 (\ln n)  + 4.4\times10^{-6} (\ln n)^2 \nonumber\\
	J_{W^\n W_R^{(n/2+1)}} &\simeq \hspace{0.8em} 0.0530 + 0.000054 (\ln n) - 3.7\times10^{-6} (\ln n)^2 .
\end{align}
$J_{W^\n W_R^{(n/2)}}+J_{W^\n W_R^{(n/2+1)}}$ is small, and
almost vanishes within numerical errors.

\subsection{Fermion loops}

\begin{figure}[t]
	\begin{minipage}{0.5 \hsize}
		\centering
		\includegraphics[bb=0 0 285 126, width=6.5cm]{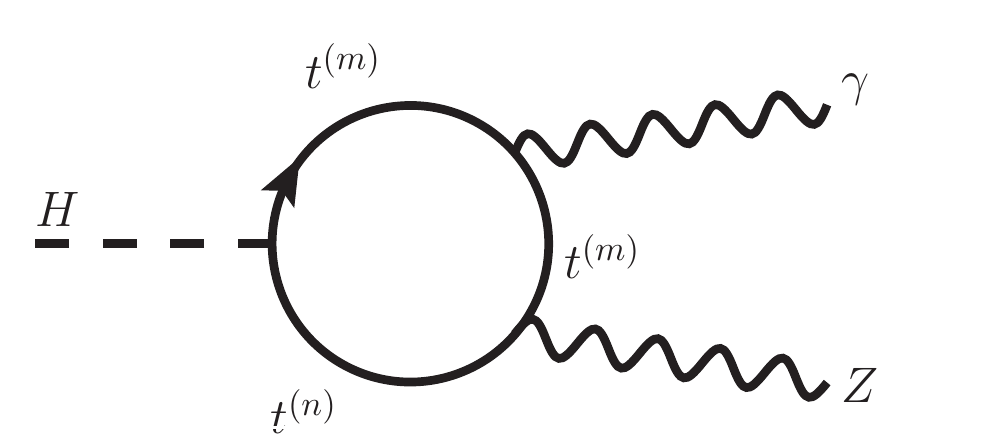}\subcaption{}
	\end{minipage}
	\begin{minipage}{0.5 \hsize}
		\centering
		\includegraphics[bb=0 0 285 126, width=6.5cm]{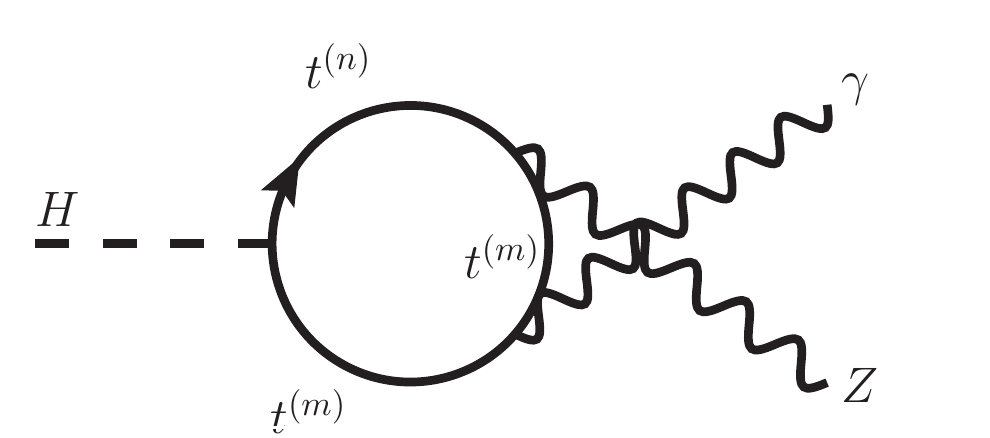}\subcaption{}
	\end{minipage}
	\begin{minipage}{0.5 \hsize}
		\centering
		\includegraphics[bb=0 0 285 126, width=6.5cm]{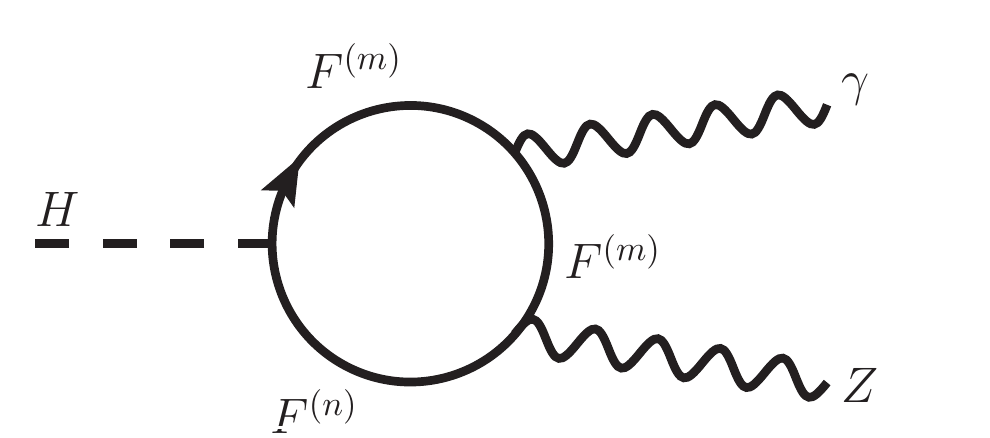}\subcaption{}
	\end{minipage}
	\begin{minipage}{0.5 \hsize}
		\centering
		\includegraphics[bb=0 0 285 126, width=6.5cm]{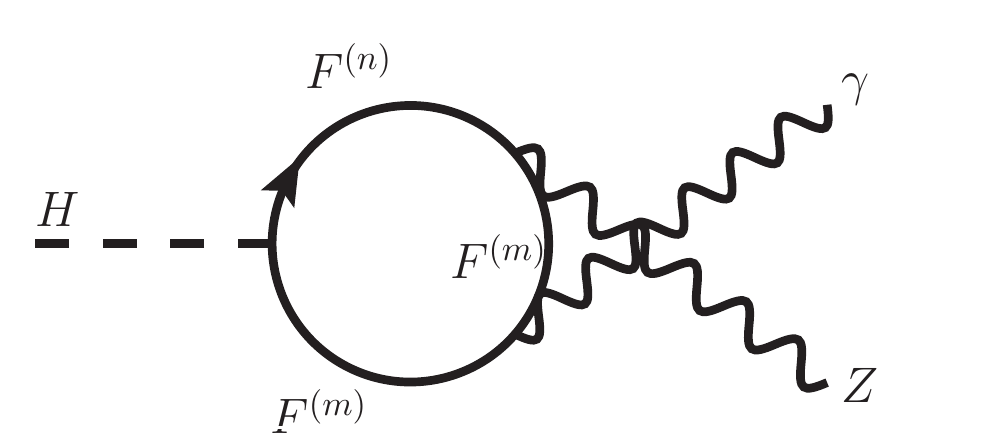}\subcaption{}
	\end{minipage}  
	\caption{The fermion loop processes of 
	$H\to Z\gamma$ decay in the $SO(5) \times U(1)$ gauge-Higgs unification. 
	$F^+$ is the charged dark fermion which does not have a zero mode. }
	\label{fermion diagrams}
\end{figure}

Contributions of fermion loops to $\Gamma (H\to Z\gamma)$ are evaluated similarly.
Top quark, charged dark fermions and their KK excitations give substantial 
contributions as shown in Fig.\ \ref{fermion diagrams}, 
whereas contributions from other quarks and leptons are negligible.

The diagrams of fermion loops~Fig.~\ref{fermion diagrams} (a, b)  (or (c,d)), 
with  generic fermions $f^{(m)}$, yield
\beqn
&&\hskip -1.cm
i\mathcal{M}^{(a)}_{f^\m,f^\n}+i\mathcal{M}^{(b)}_{f^\m,f^\n} \cr
\noalign{\kern 5pt}
&&\hskip -1.cm
   = - Q_fe \, \epsilon_\mu^* (k_1)\epsilon_\nu^* (k_2) 
 \int \frac{d^4p}{(2\pi)^4}
	 \frac{1}{p^2-m_{f^\m}^2}\frac{1}{(p-k_1)^2-m_{f^\m}^2}\frac{1}{(p-k_1-k_2)^2-m_{f^\n}^2} \cr
\noalign{\kern 5pt}
&&\hskip 0.cm 
\times \Tr \Big[
	\big(y_{f^\m f^\n}+\hat{y}_{f^\m f^\n}\gamma^5\big)(\diracslash{p}+m_{f^\m})
	 \gamma^\mu (\diracslash{p}-\diracslash{k_1}+m_{f^\m})\gamma^\nu \cr
\noalign{\kern 5pt}
&&\hskip  1.cm 
\times  \big(g^V_{Zf^\m f^\n}+g^A_{Zf^\m f^\n}\gamma^5\big)
	 (\diracslash{p}-\diracslash{k_1}-\diracslash{k_2}+m_{f^\n})\cr
\noalign{\kern 5pt}
&&\hskip 0.cm  
   +\big(y_{f^\m f^\n}+\hat{y}_{f^\n f^\m}\gamma^5\big)
	 (-\diracslash{p}+\diracslash{k_1}+\diracslash{k_2}+m_{f^\n})\gamma^\nu \cr
&&\hskip 1.cm	 
\times  \big(g^V_{Zf^\m f^\n}+g^A_{Zf^\n f^\m}\gamma^5\big)
	 (-\diracslash{p}+\diracslash{k_1}+m_{f^\m})\gamma^\mu (-\diracslash{p}+m_{f^\m})\Big].  \label{Floop1}
\eeqn
The Yukawa couplings $y_{f^\m f^\n}$ and $\hat{y}_{f^\m f^\n}$ for $f^\m = t^\m$ and $F^\m$ are 
given in \eqref{Htt3} and \eqref{HFF3}, respectively.
The amplitude \eqref{Floop1} itself involves divergent integrals, but by adding the $m\leftrightarrow n$ diagrams
and making use of $y_{f^\m f^\n}=y_{f^\n f^\m}$ and $\hat{y}_{f^\m f^\n}=-\hat{y}_{f^\n f^\m}$, 
one finds that 
\beqn
&&\hskip -1.cm
i\mathcal{M}^{(a)}_{f^\m,f^\n} +i\mathcal{M}^{(b)}_{f^\m,f^\n} + (m\leftrightarrow n )
= -\frac{iQ_fe }{4\pi^2} 
\epsilon_\mu^* (k_1)\epsilon_\nu^* (k_2)
	 \bigg(\eta^{\mu\nu}-\frac{k_2^\mu k_1^\nu}{k_1\cdot k_2}\bigg) \cr
\noalign{\kern 10pt}
&&\hskip -0.5cm
\times \Big\{ y_{f^\m f^\n}g^V_{Zf^\m f^\n}  G_+ (m_{f^\m} , m_{f^\n} )
 - \hat{y}_{f^\m f^\n}g^A_{Zf^\m f^\n}  G_- (m_{f^\m} , m_{f^\n} ) \Big\} ~,  \cr
\noalign{\kern 10pt}
&&\hskip -1.cm
G_\pm (m_1, m_2)
= 2(m_1 \pm m_2) 
+\frac{2m_Z^2(m_1 \pm m_2)}{m_H^2-m_Z^2} \big(B_0(m_H^2,m_1^2,m_2^2)
     -B_0(m_Z^2,m_1^2,m_2^2)\big)\cr
\noalign{\kern 10pt}
&&\hskip 1.5cm
+ m_1 (2m_1^2 \pm  2m_1 m_2 -m_H^2+m_Z^2)
C_0(0,m_H^2,m_Z^2,m_1^2,m_1^2,m_2^2)\ \cr
\noalign{\kern 10pt}
&&\hskip 1.5cm
\pm m_2(2m_2^2 \pm 2m_1 m_2 -m_H^2+m_Z^2)
C_0(0,m_H^2,m_Z^2,m_2^2,m_2^2,m_1^2)  ~. 
\label{Floop2}
\eeqn
In this form the amplitude is finite.

\begin{table}
\centering \footnotesize
\caption{$J_{t^\m t^\n}$ is shown for $0\le m,n\le 7$ and for $101\le m,n\le 108$ 
	in the $N_F=4$, $z_L=10^5$ case. 
	Only the values larger than $O(10^{-4})$ are shown with three significant figures.}
\label{tableJtt}
\begin{tabular}{|c|cccccccc|}\hline
    & 0 & 1 & 2 & 3 & 4 & 5 & 6 & 7 \\ \hline
  0 &    0.0988     &   -0.0041     & $O(10^{-4})$  & $O(10^{-5})$  & $O(10^{-7})$  & $O(10^{-6})$
    & $O(10^{-6})$  & $O(10^{-7})$  \\ 
  1 &   -0.0041     &   -0.0790     &    0.0638	    & $O(10^{-5})$  & $O(10^{-4})$  & $O(10^{-9})$  
    & $O(10^{-10})$ & $O(10^{-8})$  \\
  2 & $O(10^{-4})$  &    0.0638     &   -0.0350     &   -0.0071     & $O(10^{-6})$  & $O(10^{-6})$
    & $O(10^{-9})$  & $O(10^{-5})$  \\
  3 & $O(10^{-5})$  & $O(10^{-5})$  &   -0.0071     &   -0.0763     &    0.0616     & $O(10^{-6})$
    & $O(10^{-4})$  & $O(10^{-9})$  \\
  4 & $O(10^{-7})$  & $O(10^{-4})$  & $O(10^{-6})$  &    0.0616     &   -0.0338     &   -0.0071  
    & $O(10^{-6})$  & $O(10^{-6})$  \\
  5 & $O(10^{-6})$  & $O(10^{-9})$  & $O(10^{-6})$  & $O(10^{-6})$  &   -0.0071     &   -0.0754 
    &    0.0609     & $O(10^{-6})$  \\
  6 & $O(10^{-6})$  & $O(10^{-10})$ & $O(10^{-9})$  & $O(10^{-4})$  & $O(10^{-6})$  &    0.0609
    &   -0.0334     &   -0.0070     \\
  7 & $O(10^{-7})$  & $O(10^{-8})$  & $O(10^{-5})$  & $O(10^{-9})$  & $O(10^{-6})$  & $O(10^{-6})$
    &   -0.0070     &   -0.0751     \\
\hline\end{tabular}
\vskip 6pt
\begin{tabular}{|c|cccccccc|}\hline
    & 101 & 102 & 103 & 104 & 105 & 106 & 107 & 108 \\ \hline
101 &   -0.0761     &    0.0610     & $O(10^{-6})$  & $O(10^{-4})$  & $O(10^{-13})$ 
    & $O(10^{-7})$  & $O(10^{-8})$  & $O(10^{-6})$  \\
102 &    0.0610     &   -0.0337     &   -0.0068     & $O(10^{-6})$  & $O(10^{-6})$
    & $O(10^{-13})$ & $O(10^{-5})$  & $O(10^{-9})$  \\
103 & $O(10^{-6})$  &   -0.0068     &   -0.0761     &    0.0610     & $O(10^{-6})$
    & $O(10^{-4})$  & $O(10^{-13})$ & $O(10^{-7})$  \\
104 & $O(10^{-4})$  & $O(10^{-6})$  &    0.0610     &   -0.0337     &   -0.0068  
    & $O(10^{-6})$  & $O(10^{-6})$  & $O(10^{-13})$ \\
105 & $O(10^{-13})$ & $O(10^{-6})$  & $O(10^{-6})$  &  -0.0068      &   -0.0761
    &    0.0610     & $O(10^{-6})$  & $O(10^{-4})$  \\
106 & $O(10^{-7})$  & $O(10^{-13})$ & $O(10^{-4})$  & $O(10^{-6})$  &    0.0610
    &   -0.0337     &   -0.0068     & $O(10^{-6})$  \\
107 & $O(10^{-8})$  & $O(10^{-5})$  & $O(10^{-13})$ & $O(10^{-6})$  & $O(10^{-6})$
    &   -0.0068     &   -0.0762     &    0.0610     \\
108 & $O(10^{-6})$  & $O(10^{-9})$  & $O(10^{-7})$  & $O(10^{-13})$ & $O(10^{-4})$ 
    & $O(10^{-6})$  &    0.0610     &   -0.0337     \\
\hline\end{tabular}\end{table}

\begin{table}
\centering \footnotesize
\caption{$J_{F^{+(m)} F^{+(n)}}$ is shown for $0\le m,n\le 7$ and for $101\le m,n\le 108$
 in the $N_F=4$, $z_L=10^5$ case. 
Only the values larger than $O(10^{-4})$ are shown with three significant figures.}
\label{tableJFF}
\begin{tabular}{|c|ccccccc|}\hline
    & 1 & 2 & 3 & 4 & 5 & 6 & 7  \\ \hline
  1 &    0.2256    &   -0.0272    & $O(10^{-5})$ &   -0.0040    & $O(10^{-8})$ 
    & $O(10^{-5})$ & $O(10^{-7})$ \\
  2 &   -0.0272    &    0.2378    &    0.0824    & $O(10^{-6})$ & $O(10^{-5})$
    & $O(10^{-8})$ & $O(10^{-5})$ \\
  3 & $O(10^{-5})$ &    0.0824    &    0.2204    &   -0.3188    & $O(10^{-6})$
    &    0.0036    & $O(10^{-8})$ \\
  4 &   -0.0040    & $O(10^{-6})$ &   -0.3188    &    0.2554    &    0.0866   
    & $O(10^{-6})$ & $O(10^{-5})$ \\
  5 & $O(10^{-8})$ & $O(10^{-5})$ & $O(10^{-6})$ &    0.0866    &    0.2245
    &   -0.3263    & $O(10^{-6})$ \\
  6 & $O(10^{-5})$ & $O(10^{-8})$ &   -0.0036    & $O(10^{-6})$ &   -0.3263
    &    0.2612    &    0.0874    \\
  7 & $O(10^{-7})$ & $O(10^{-5})$ & $O(10^{-8})$ & $O(10^{-5})$ & $O(10^{-6})$
    &    0.0874    &    0.2271    \\
\hline\end{tabular}
\vskip 6pt
\begin{tabular}{|c|cccccccc|}\hline
    & 101 & 102 & 103 & 104 & 105 & 106 & 107 & 108 \\ \hline
101 &    0.2505     &   -0.3528     & $O(10^{-6})$  &   -0.0033     & $O(10^{-11})$ 
    & $O(10^{-5})$  & $O(10^{-8})$  & $O(10^{-5})$  \\
102 &   -0.3528     &    0.2918     &    0.0848     & $O(10^{-5})$  & $O(10^{-5})$
    & $O(10^{-12})$ & $O(10^{-5})$  & $O(10^{-8})$  \\
103 & $O(10^{-6})$  &    0.0848     &    0.2508     &   -0.3531     & $O(10^{-6})$
    &   -0.0033     & $O(10^{-11})$ & $O(10^{-5})$  \\
104 &   -0.0033     & $O(10^{-5})$  &   -0.3530     &    0.2921     &    0.0848   
    & $O(10^{-5})$  & $O(10^{-5})$  & $O(10^{-12})$ \\
105 & $O(10^{-11})$ & $O(10^{-5})$  & $O(10^{-6})$  &    0.0848     &    0.2510
    &   -0.3533     & $O(10^{-6})$  &   -0.0033     \\
106 & $O(10^{-5})$  & $O(10^{-12})$ &   -0.0033     & $O(10^{-5})$  &   -0.3533
    &    0.2924     &    0.0847     & $O(10^{-5})$  \\
107 & $O(10^{-8})$  & $O(10^{-5})$  & $O(10^{-11})$ & $O(10^{-5})$  & $O(10^{-6})$
    &    0.0847     &    0.2512     &   -0.3535     \\
108 & $O(10^{-5})$  & $O(10^{-8})$  & $O(10^{-5})$  & $O(10^{-12})$ &   -0.0033 
    & $O(10^{-5})$  &   -0.3535     &    0.2927     \\
\hline\end{tabular}\end{table}

We define 
$J_{t^\m t^\n}$ and $J_{F^{+(m)} F^{+(n)}}$ by
\begin{align}
	J_{t^\m t^\n}&\equiv \frac{g^V_{Zt^\m t^\n}y_{t^\m t^\n}\cos\theta_W }{g_w y_t\cos\theta_H}, \\
	J_{F^{+(m)} F^{+(n)}}&\equiv 
	\frac{g^V_{ZF^{+(m)} F^{+(n)}}y_{F^\m F^\n}\cos\theta_W }{g_w y_t\sin\frac{\theta_H}{2}}
\end{align}
In the Table~\ref{tableJtt} and Table~\ref{tableJFF}, $J_{t^\m t^\n}$ and $J_{F^{+(m)} F^{+(n)}}$ 
by the numerical evaluation are shown. 
As in the case of   $J_{W^\n W^\n}$,  $J_{t^\m t^\n}$ and $J_{F^{+(m)} F^{+(n)}}$ for $|m-n|\ge 2$ 
become negligible for $m,n > 100$. 
In addition, the terms proportional to $\hat{y}_{f^\m f^\n} g^A_{Zf^\m f^\n}$ are negligible  around $m,n=100$.
The ratio  $(\hat{y}_{f^\m f^\n}/y_{f^\m f^\n}) \cdot(g^A_{Zf^\m f^\n}/g^V_{Zf^\m f^\n})$ 
is smaller than $10^{-4}$, and the $\hat{y}_{f^\m f^\n} g^A_{Zf^\m f^\n}$ term in \eqref{Floop2} may be
dropped.

For the asymptotic behavior of the amplitude for $m,n \gg 1$ only the $|m-n|\le 1$ terms are relevant. 
For $|m-n|\le 1$,  $m_{f^{(n\pm 1)}}\simeq m_{f^\n}$ and the amplitudes are evaluated as
\beqn
&&\hskip -1.cm
i\mathcal{M}_\text{fermion}\cr
&&\hskip -1.cm
\equiv\frac{1}{2}\sum_{m,n}^\infty \Big\{i\mathcal{M}^{(a)}_{t^\m,t^\n} +i\mathcal{M}^{(b)}_{t^\m,t^\n} + 
i\mathcal{M}^{(c)}_{F^{+(m)},F^{+(n)}}+i\mathcal{M}^{(d)}_{F^{+(m)},F^{+(n)}}
+ (m \leftrightarrow n) \Big\} \cr
\noalign{\kern 10pt}
&&\hskip -1.cm
\approx
\sum_n^\infty\frac{i}{4\pi^2}\epsilon_\mu^* (k_1)\epsilon_\nu^* (k_2)
\bigg(\eta^{\mu\nu}-\frac{k_2^\mu k_1^\nu}{k_1\cdot k_2}\bigg)
\frac{ e g_w y_t\cos\theta_H}{\cos\theta_W}  (m_H^2-m_Z^2) \cr
\noalign{\kern 10pt}
&&\hskip -1.cm
\times \bigg\{ \frac{2}{3m_{t^\n}} 
(J_{t^\n t^\n}+J_{t^{(n+1)} t^\n}+J_{t^{(n-1)} t^\n})
	\Big[ I_1(\tau_{t^\n},\lambda_{t^\n})-I_2(\tau_{t^\n},\lambda_{t^\n}) \Big]  \cr
\noalign{\kern 10pt}
&&\hskip -.5cm
+	\frac{1}{m_{F^{+(n)}}} 
(J_{F^{+(n)} F^{+(n)}}+J_{F^{+(n+1)} F^{+(n)}}+J_{F^{+(n-1)} F^{+(n)}}) \cr
\noalign{\kern 10pt}
&&\hskip +3.cm
\times \Big[ I_1(\tau_{F^{+(n)}},\lambda_{F^{+(n)}})-I_2(\tau_{F^{+(n)}},\lambda_{F^{+(n)}}) \Big] 
\bigg\}~.
\label{Floop3}
\eeqn
\\

\begin{figure}[thb]
\centering
\includegraphics[bb=0 0 415 190, width=12cm]{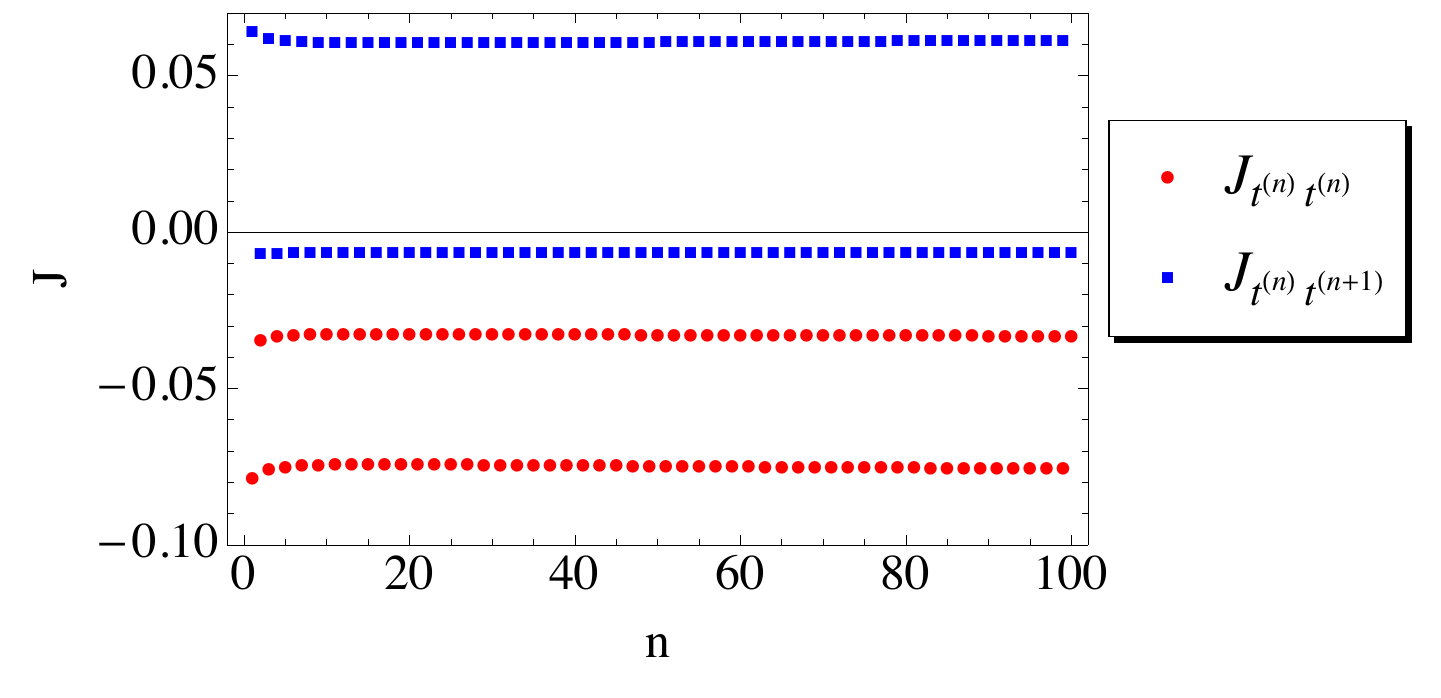}
\caption{$J_{t^\n t^\n}$ and $J_{t^\n t^{(n+1)}}$ are plotted for $1\le n \le 100$ in the $N_F=4$, $z_L=10^5$ case. 
The red circles and blue squares express  $J_{t^\n t^\n}$ and $J_{t^\n t^{(n+1)}}$, respectively.} 
	\label{Jtt-plot}
\end{figure}
\begin{figure}[thb]
\centering
\includegraphics[bb=0 0 415 201, width=13cm]{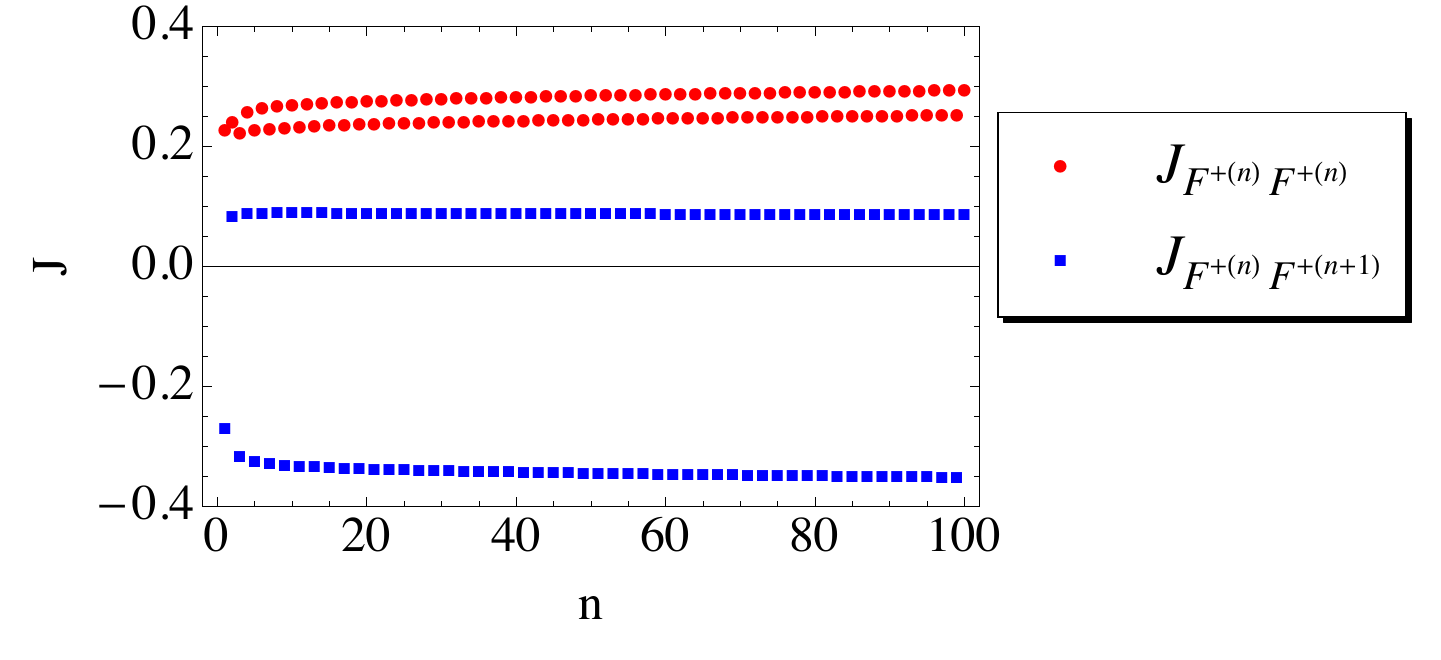}
\caption{$J_{F^{+(n)} F^{+(n)}}$ and $J_{F^{+(n)} F^{+(n+1)}}$ are plotted for $1\le n \le 100$ in the $N_F=4$, $z_L=10^5$ case. The red circles and blue squares express  $J_{F^{+(n)}F^{+(n)}}$ and $J_{F^{+(n)}F^{+(n+1)}}$,
 respectively.} 
\label{JFF-plot}
\end{figure}

$J_{t^\n t^\m}$ and $J_{F^{+(n)} F^{+(m)}}$ ($m=n, n+1$) are plotted in 
Figs.\ \ref{Jtt-plot} and \ref{JFF-plot}.
They are approximately given, for $101 \le n \le 200$, by
\begin{align}
&\hskip -1.cm
J_{t^\n t^\n}   \simeq -0.0597+0.00323(\ln n)-0.00047(\ln n)^2 \nonumber\\
&\hskip 1.cm
+(-1)^{n-1} \big\{ -0.0230+0.00122(\ln n)-0.00018(\ln n)^2 \big\} , \nonumber\\
\noalign{\kern 5pt}
&\hskip -1.cm
J_{t^\n t^{(n+1)}}  \simeq  0.0296-0.00162(\ln n)+0.00024(\ln n)^2  \nonumber\\
\noalign{\kern 5pt}
&\hskip 1.cm
+(-1)^{n-1} \big\{   0.0362-0.00143(\ln n)+0.00020(\ln n)^2 \big\}, \\
\noalign{\kern 5pt}
&\hskip -1.cm 
J_{F^{+(n)} F^{+(n)}}   \simeq  0.280-0.0172(\ln n)+0.00331(\ln n)^2 \nonumber\\
&\hskip 1.cm
+(-1)^{n-1} \big\{ -0.0212+0.00131(\ln n)-0.00026(\ln n)^2 \big\} ,  \nonumber\\
&\hskip -1.cm
J_{F^{+(n)}F^{+(n+1)}}  \simeq -0.138+0.0084(\ln n)-0.00163(\ln n)^2 \nonumber\\
&\hskip 1.cm
+(-1)^{n-1}\big\{ -0.2218+0.00558(\ln n)-0.00107(\ln n)^2 \big\} .
\label{Jtt2}
\end{align}
Adding them, one finds, within numerical errors,  that
\beqn
&&\hskip -1.cm
J_{t^\n t^\n}+J_{t^{(n+1)}t^\n }+J_{t^{(n-1)} t^\n} \cr
\noalign{\kern 5pt}
&&\hskip -.5cm
\approx (-1)^{n-1} \big\{ -0.0230+0.00122(\ln n)-0.00018(\ln n)^2 \big\} , \cr
\noalign{\kern 5pt}
&&\hskip -1.cm
J_{F^{+(n)} F^{+(n)}}+J_{F^{+(n+1)}F^{+(n)}}+J_{F^{+(n-1)}F^{+(n)}} \cr
\noalign{\kern 5pt}
&&\hskip -.5cm
\approx (-1)^{n-1} \big\{ -0.0212+0.00131(\ln n)-0.00026(\ln n)^2 \big\} .
\label{Jtt3}
\eeqn
Since $i\mathcal{M}_f \propto \sum_n (J_{f^\n f^\n}+J_{f^{(n+1)} f^\n}+J_{f^{(n-1)} f^\n})/n$
for large $n$,  the sum converges as in the case of the gauge boson loops.

\subsection{Total amplitude}

The ratio of the sum  of the all boson contributions to the $W$ boson contribution
and that of all the fermion contributions to the top quark contribution are given by
\begin{align}
	\frac{\mathcal{M}_\text{boson}}{\mathcal{M}_{W^{(0)}\text{only}}} = 0.9998 , \quad
	\frac{\mathcal{M}_\text{fermion}}{\mathcal{M}_{t^{(0)}\text{only}}} = 1.0023, 
\end{align}
for $z_L = 10^5$, $n_F=4$, and $\theta_H = 0.1153$, respectively 
where $\mathcal{M}_\text{boson}\equiv\mathcal{M}_W+\mathcal{M}_{W_R}$. 
The ratio of the whole amplitude to the $W$ boson and top quark contributions  is
\begin{align}
\frac{\mathcal{M}_\text{boson}+\mathcal{M}_\text{fermion}}{\mathcal{M}_{W^{(0)}\text{only}}
+\mathcal{M}_{t^{(0)}\text{only}}} = 0.9997, 
\end{align}
One finds that the KK mode contributions are negligible.
As  $g_{HW^{(0)}W^{(0)}} \simeq g_w \cos\theta_H$ and $y_{t^{(0)}}  \simeq y_{t \text{SM}}\cos\theta_H$,
the decay width is approximated by
\begin{align}
	\Gamma(H\to Z\gamma)_\text{GHU}\simeq \Gamma(H\to Z\gamma)_\text{SM} \times \cos^2\theta_H .
\end{align}

In the gauge-Higgs unification,  the decay width of $H\to WW$, $H\to ZZ$, $H\to bb$ and $H\to \tau\tau$ are 
suppressed by $\cos^2\theta_H$ at the tree level.
The decay width of the $H\to \gamma\gamma$ and $H\to Z\gamma$ are also suppressed by $\cos^2\theta_H$.
Therefore the branching ratios of the Higgs decay modes in this model are almost the same as in the SM. 
The dominant process in the Higgs boson production is $gg\to H$, 
and the production cross section  is also suppressed by $\cos^2\theta_H$.
Therefore the signal strength, 
$\sigma(gg\to H)  B(H\to Z\gamma)/ [\sigma(gg\to H)   B(H\to Z\gamma) ]_\SM$, is 
approximately $\cos^2\theta_H$ as in the other decay modes.
For $\theta_H \sim 0.1$, the deviation from the SM amounts only 1\%.

We stress the finiteness of $\Gamma(H\to Z\gamma)$ in the gauge-Higgs unification
which results from non-trivial cancellation among contributions of the KK modes.
One might wonder why such cancellation takes place and what underlies the finiteness
of the amplitude ${\cal M} (H\to Z\gamma)$ in the gauge-Higgs unification.  
We argue that it is guaranteed by the gauge invariance and by the fact that the 4D Higgs field 
$H$ is the fluctuation mode of the AB phase $\theta_H$.
In the effective action  the 4D Higgs field $H(x)$ and  AB phase $\theta_H$  appear 
in the combination of $\theta_H + H(x)/f_H$ so that 
${\cal M} (H\to Z\gamma)$ is related to $(\dd/\dd \theta_H) \Pi_{Z \gamma} (\theta_H)$
where $\Pi_{Z \gamma} (\theta_H)$ is the $Z \gamma$ vacuum polarization.
The 5D gauge invariance guarantees that $\Pi_{Z \gamma} (\theta_H)$  is periodic in $\theta_H$, 
and can be expanded in a Fourier series 
$\Pi_{Z \gamma} (\theta_H) = \sum_n  \alpha_n e^{in\theta_H}$.
The $\theta_H$-dependence of the $Z$ couplings of the fields running the inside loops
is known to be very weak.  At the one loop level the dominant $\theta_H$-dependence of
$\Pi_{Z \gamma} (\theta_H)$ comes from the propagators inside the loop.  Hence the divergence
degree is lowered by differentiating $\Pi_{Z \gamma} (\theta_H)$ with respect to $\theta_H$.
There should exist a positive integer $q$ such that 
$\dd^q \Pi_{Z \gamma}/ \dd \theta_H^q = \sum_{n \not= 0} (in)^q \alpha_n e^{in\theta_H}$
is finite.  This in turn implies that $\Pi_{Z \gamma} (\theta_H) - \alpha_0$ is finite, from which
the finiteness of ${\cal M} (H\to Z\gamma)$ at the one loop level follows.  A similar argument
has been employed to prove the finiteness of the effective potential $V_\eff (\theta_H)$
at the one loop level.\cite{Hosotani-2005-rev}

\section{Conclusion and discussions}

In this paper we have evaluated the decay rate  $\Gamma ( H\to Z\gamma)$ in the $SO(5)\times U(1)$ 
gauge-Higgs unification.  The processes $H\to \gamma \gamma$ and $H\to Z\gamma$ 
do not occur at the tree level.  They do proceed  at the one loop level, where 
an infinite number of the KK modes of gauge bosons and fermions give loop corrections. 
Contrary to the naive expectation that the sum of an infinite number of the KK mode contributions
may yield substantial corrections to the decay rates in the SM, it has been known that 
the correction to $\Gamma (H\to \gamma \gamma)$ coming from KK modes turns
out very tiny, thanks to the cancellation among the KK mode corrections. 

We have  examined  the process  $H\to Z\gamma$ in detail, for which the KK number need not be
conserved  inside the loop.  We have shown, by direct evaluation, that there appears miraculous 
cancellation among the loop diagrams in which the KK number is conserved in the loop and those
in which the KK number is not conserved.  As a result  the amplitude for   $H\to Z\gamma$ becomes
finite.  We also showed that the correction due to various KK modes is very small.
$\Gamma ( H\to Z\gamma)$ in the gauge-Higgs unification is approximately 
$\cos^2 \theta_H$ times that in the SM. The deviation from the SM is very small for $\theta_H < 0.1$.

The result is very promising.  The $SO(5) \times U(1)$ gauge-Higgs unification yields almost 
the same phenomenology as the SM at low energies.  At higher energy scale it predicts
KK excitation modes as $Z'$ and $W'$ events and a dark matter candidate, 
which awaits confirmation at 14 TeV LHC and by DM direct-detection experiments. 
Small deviations of $HWW$ and $HZZ$ couplings from the SM  will be checked
in future colliders.\cite{Asner:2013psa}
In addition to deriving more predictions for collider experiments, the scenario of the gauge-Higgs
unification has to be refined.  The scenario of the gauge-Higgs grand unification has been 
proposed.\cite{Hosotani:2015hoa}
An attempt has been made to dynamically determine orbifold boundary conditions.\cite{Yamamoto:2013oja}
The Hosotani mechanism, essential for the electroweak gauge symmetry breaking 
in the gauge-Higgs unification, has been investigated nonperturbatively on the lattice.\cite{Cossu:2013ora}
The gauge-Higgs unification is one of the keys to investigate the extra dimension.

\vskip .5cm

\subsection*{Acknowledgements}
This work was supported in part by Japan Society for the Promotion of Science, Grants-in-Aid for Scientific Research, 
No.\ 23104009 (YH),  No.\ 21244036 (YH),
and National Research Foundation of Korea, 2012R1A2A1A01006053 (HH).

\appendix

\section{Couplings of KK modes to $Z$ and $H$}
We summarize the $Z$ and $H$ couplings of the KK modes relevant
for $H \go Z \gamma$.

\subsection{Base functions}
Mode functions for KK towers of various fields in the RS spacetime are expressed 
in terms of Bessel functions.  
For gauge fields we define
\beqn
&&\hskip -1cm 
C(z;\lambda) = \frac{\pi}{2}\lambda   z z_L F_{1,0}(\lambda z, \lambda z_L) ~, \quad
C'(z;\lambda)= \frac{\pi}{2}\lambda^2 z z_L F_{0,0}(\lambda z, \lambda z_L) ~, \cr \noalign{\kern 5pt}
&&\hskip -1cm 
S(z;\lambda) = -\frac{\pi}{2}\lambda   z  F_{1,1}(\lambda z, \lambda z_L) ~, \quad
S'(z;\lambda)= -\frac{\pi}{2}\lambda^2 z F_{0,1}(\lambda z,  \lambda z_L) ~, \cr \noalign{\kern 5pt}
&&\hskip -1cm 
\hat S(z;\lambda) = \frac{C(1; \lambda)}{S(1; \lambda)} \, S(z; \lambda) ~,  
\label{BesselF1}
\eeqn
where $F_{\alpha,\beta}(u,v) = J_\alpha (u) Y_\beta(v)-Y_\alpha(u) J_\beta(v)$.  
They satisfy
\beqn
&&\hskip -1cm 
C(z_L; \lambda) = z_L ~,~~ C' (z_L; \lambda) =0~, ~~ 
S(z_L; \lambda) = 0 ~,~~ S' (z_L; \lambda) =\lambda ~,  \cr \noalign{\kern 5pt}
&&\hskip -1cm 
C S' - S C' = \lambda z ~. \label{BesselF2}
\eeqn
For fermions with a bulk mass parameter $c$ we define
\beqn
&&\hskip -1cm
\begin{pmatrix} C_L \cr S_L \end{pmatrix}  (z;\lambda, c)
= \pm \frac{\pi}{2} \lambda\sqrt{zz_L} F_{c+{1\over 2},c\mp{1\over 2}}  (\lambda z, \lambda z_L) ~, \cr
\noalign{\kern 10pt}
&&\hskip -1cm  
\begin{pmatrix} C_R \cr S_R \end{pmatrix}  (z;\lambda, c)
= \mp \frac{\pi}{2} \lambda\sqrt{zz_L} F_{c-{1\over 2},c\pm {1\over 2}} (\lambda z, \lambda z_L)~. \label{BesselF3}
\eeqn
They satisfy
\beqn
&&\hskip -1.cm
D_+ (c)\begin{pmatrix} C_L \cr S_L \end{pmatrix} = \lambda \begin{pmatrix} S_R \cr C_R \end{pmatrix} ~,~~
D_- (c)\begin{pmatrix} C_R \cr S_R \end{pmatrix} = \lambda \begin{pmatrix} S_L \cr C_L \end{pmatrix} ~, \cr
\noalign{\kern 5pt}
&&\hskip 1.cm
D_\pm (c)  = \pm \frac{d}{dz} + \frac{c}{z}~, \cr
\noalign{\kern 5pt}
&&\hskip -1.cm
C_R=C_L =1 ~,~~  S_R=S_L=0 ~,   ~~{\rm at~} z= z_L ~, \cr 
\noalign{\kern 5pt}
&&\hskip -1.cm
C_L C_R - S_L S_R =1 ~.
\label{BesselF4}
\eeqn

In the following we evaluate various couplings by inserting the formulas for the KK expansions.
Basic formulas for the KK expansions of gauge fields and quark fields are summarized 
in Ref.~\cite{Funatsu:2014fda}, whereas those for dark fermions are given in Ref.~\cite{Funatsu:2014tka}.
We adopt the same notation as in those references.
The numerical values for the various couplings are given for the parameter set
\begin{equation}
z_L=10^5, \theta_H=0.1153, m_\text{KK}=7.405 \text{ TeV}, k= 2.357\times10^8\text{ GeV}, 
c_t=0.2270, c_F=0.3321 .
\end{equation}

\subsection{$ZW^\m W^\n$ coupling}
The $ZW^\m W^\n$ coupling is contained in 
\begin{align}
&\int_1^{z_L} \frac{dz}{kz}\left(-\frac{1}{4}\right)
\text{Tr}\left[F_{\mu\nu}F_{\rho\sigma}\right]
\eta^{\mu \rho}\eta^{\nu\sigma} \cr
&\supset ig_A\int_1^{z_L}\frac{dz}{kz}
\text{Tr} \Big[
 (\partial_\mu \hat{Z}_\nu-\partial_\nu \hat{Z}_\mu)[\hat{W}^+_\rho,\hat{W}^-_\sigma] \cr & \qquad
+(\partial_\mu \hat{W}^-_\nu-\partial_\nu \hat{W}^-_\mu)[\hat{Z}_\rho,\hat{W}^+_\sigma] 
+(\partial_\mu \hat{W}^+_\nu-\partial_\nu \hat{W}^+_\mu)[\hat{Z}_\rho,\hat{W}^-_\sigma]\Big] 
\eta^{\mu \rho}\eta^{\nu\sigma}\nonumber\\
&\supset i\sum_{m,n}g_{ZW^\m W^\n}\eta^{\mu \rho}\eta^{\nu\sigma}
\Big\{(\partial_\mu Z_\nu -\partial_\nu Z_\mu) W^{+(m)}_\rho W^{-(n)}_\sigma \nonumber\\
&-(\partial_\mu W^{+(m)}_\nu -\partial_\nu W^{+(m)}_\mu) Z_\rho W^{-(n)}_\sigma +(\partial_\mu W^{-(n)}_\nu -\partial_\nu W^{-(n)}_\mu) Z_\rho W^{+(m)}_\sigma \Big\}
\end{align}
so that one finds that
\beqn
&&\hskip -1.cm
g_{ZW^\m W^\n} = g_w \sqrt{L} \int_1^{z_L} \frac{dz}{kz}\cr
\noalign{\kern 5pt}
&&\hskip -.5cm 
\times\biggl\{ h_Z^L \bigg( h_{W^\m}^Lh_{W^\n}^L+  \frac{\hat{h}_{W^\m}\hat{h}_{W^\n}}{2}\bigg)+h_Z^R \bigg( h_{W^\m}^Rh_{W^\n}^R+\frac{\hat{h}_{W^\m}\hat{h}_{W^\n}}{2}\bigg) \cr
\noalign{\kern 5pt}
&&\hskip -.0cm 
+\hat{h}_Z\bigg( \frac{h_{W^\m}^L\hat{h}_{W^\n}+  h^R_{W^\m}\hat{h}_{W^\n}
+\hat{h}_{W^\m}h_{W^\n}^L+  \hat{h}_{W^\m}h_{W^\n}^R}{2}\bigg)\bigg\}  \cr
\noalign{\kern 5pt}
&&\hskip -1.0cm 
= g_w \cos\theta_W \frac{\sqrt{L}}{\sqrt{r_Z\:r_{W^\m}\:r_{W^\n}}}
\int_1^{z_L} \frac{dz}{kz}\frac{1}{\sqrt{2}} \cr
\noalign{\kern 5pt}
&&\hskip -.5cm 
\times\bigg[  2C_Z \big((1+\cos^2\theta_H)C_{W^\m}C_{W^\n}
+\sin^2\theta_H\hat{S}_{W^\m}\hat{S}_{W^\n}\big) \cr
\noalign{\kern 5pt}
&&\hskip -1.cm 
+\frac{\sin^2\theta_H}{\cos^2\theta_W}
\Big\{-C_Z(C_{W^\m}C_{W^\n}+\hat{S}_{W^\m}\hat{S}_{W^\n})
+\hat{S}_Z(C_{W^\m}\hat{S}_{W^\n}+\hat{S}_{W^\m}C_{W^\n})    \Big\}\bigg] .
\label{ZWW1}
\eeqn
Here $C_{W^{(m)}} = C(z; \lambda_{W^{(m)}})$ etc.
Numerical values of  $g_{ZW^\m W^\n}$ are given in Table \ref{tbl:ZWW}.

\begin{table}[thbp]\centering \footnotesize
\caption{$g_{ZW^\m W^\n}/g_w\cos\theta_W$. Only the values larger than $O(10^{-3})$ are shown and written by three significant figures.}\label{tbl:ZWW}
\begin{tabular}{|c|cccccccc|}\hline
   &0&1&2&3&4&5&6&7\\ \hline
 0 &     1.       & $O(10^{-4})$ & $O(10^{-7})$ & $O(10^{-5})$ & $O(10^{-7})$ & $O(10^{-6})$    & $O(10^{-8})$ & $O(10^{-6})$ \\
 1 & $O(10^{-4})$ &    0.996     &    0.032     & $O(10^{-5})$ & $O(10^{-4})$ & $O(10^{-6})$    & $O(10^{-5})$ & $O(10^{-6})$ \\
 2 & $O(10^{-7})$ &    0.032     &    0.350     &   -0.022     & $O(10^{-7})$ & $O(10^{-4})$    & $O(10^{-6})$ & $O(10^{-4})$ \\
 3 & $O(10^{-5})$ & $O(10^{-5})$ &   -0.022     &    0.996     &    0.032     & $O(10^{-5})$    & $O(10^{-4})$ & $O(10^{-6})$ \\
 4 & $O(10^{-7})$ & $O(10^{-4})$ & $O(10^{-7})$ &    0.032     &    0.350     &   -0.023     & $O(10^{-5})$ & $O(10^{-4})$ \\
 5 & $O(10^{-6})$ & $O(10^{-6})$ & $O(10^{-4})$ & $O(10^{-5})$ &   -0.023     &    0.996     &    0.032     & $O(10^{-5})$ \\
 6 & $O(10^{-8})$ & $O(10^{-5})$ & $O(10^{-6})$ & $O(10^{-4})$ & $O(10^{-5})$ &    0.032     &    0.350     &   -0.023     \\
 7 & $O(10^{-6})$ & $O(10^{-6})$ & $O(10^{-4})$ & $O(10^{-6})$ & $O(10^{-4})$ & $O(10^{-5})$ &   -0.023     &    0.996     \\
\hline 
\end{tabular}\end{table}

\subsection{$\gamma ZW^\m W^\n$ coupling}
Similarly $\gamma ZW^\m W^\n$ coupling is contained in 
\beqn
&&\hskip -1.cm
(g_A)^2\int_1^{z_L}\frac{dz}{kz}
\text{Tr}\Big[[\hat{A}^{\gamma(A)}_\mu, \hat{W}^+_\nu]
\left([\hat{Z}_\rho, \hat{W}^-_\sigma]-[\hat{Z}_\sigma, \hat{W}^-_\rho]\right) \nonumber\\
&&\hskip 2.cm+
[\hat{A}^{\gamma(A)}_\mu, \hat{W}^-_\nu]
\left([\hat{Z}_\rho, \hat{W}^+_\sigma]-[\hat{Z}_\sigma, \hat{W}^+_\rho]\right)\Big]
\eta^{\mu \rho}\eta^{\nu\sigma} \cr
\noalign{\kern 5pt}
&& \hskip -1.cm
\supset \sum_{m,n}g_{\gamma ZW^\m W^\n}\eta^{\mu \rho}\eta^{\nu\sigma} \nonumber\\
&&\times\{
 A^{\gamma}_\mu W^{+(m)}_\nu (Z_\rho W^{-(n)}_\sigma - Z_\sigma W^{-(n)}_\rho)
+A^{\gamma}_\mu W^{-(n)}_\nu (Z_\rho W^{+(m)}_\sigma - Z_\sigma W^{+(m)}_\rho)\} 
\eeqn
so that
\beqn
&&\hskip -1.cm
g_{\gamma ZW^\m W^\n}= -g_w^2L\int_1^{z_L} \frac{dz}{kz} \cr \noalign{\kern 5pt}
&&\hskip -.5cm
\times\biggl\{ 
 h_\gamma^L h_{W^\m}^L \bigg( h_Z^Lh_{W^\n}^L+ \frac{\hat{h}_Z\hat{h}_{W^\n}}{2}\bigg)
+h_\gamma^R h_{W^\m}^R \bigg( h_Z^Rh_{W^\n}^R+ \frac{\hat{h}_Z\hat{h}_{W^\n}}{2}\bigg) \cr
\noalign{\kern 5pt}
&&\hskip -.0cm
+\frac{h_\gamma^L+h_\gamma^R}{2}\hat{h}_{W^\m}
\bigg(\frac{h_Z^L\hat{h}_{W^\n}+h^R_Z\hat{h}_{W^\n}+\hat{h}_Z h_{W^\n}^L+\hat{h}_Zh_{W^\n}^R}{2}
\bigg)\bigg\} \cr
\noalign{\kern 5pt}
&&\hskip -1.cm
= -e g_{ZW^\m W^\n} ~.
\label{gammaZWW1}
\eeqn
The relation $g_{\gamma ZW^\m W^\n}= -e g_{ZW^\m W^\n}$ follows from the gauge invariance as well.

\subsection{$HW^\m W^\n$ coupling}
The Higgs coupling $HW^\m W^\n$ is contained in the $\Tr F_{\mu z} F^{\mu z} $ term 
\begin{align}
&-i g_A k^2 \int_1^{z_L}\frac{dz}{kz} 
\text{Tr}\left[\partial_z \hat{W}^-_\mu \big[\hat{W}^+_\nu, \hat{H}\big]+
\partial_z \hat{W}^+_\mu \big[\hat{W}^-_\nu, \hat{H}\big]\right]\eta^{\mu\nu}\nonumber\\
&\supset -\sum_{m,n}g_{HW^\m W^\n} H W^{+(m)}_\mu W^{-(n)}_\nu\; \eta^{\mu\nu}
\end{align}
so that
\beqn
&&\hskip -1.cm
g_{HW^\m W^\n}
=i g_A k^2 \int_1^{z_L}\frac{dz}{kz} \cr
\noalign{\kern 5pt}
&&\hskip -1.cm
\times \frac{i}{2}u_H(z)
\left[-\big(\partial_z \hat{h}_{W^\m}\big)\big(h^L_{W^\n}-h^R_{W^\n}\big)
+\partial_z\big(h^L_{W^\m}-h^R_{W^\m}\big)\hat{h}_{W^\n}
+(m \longleftrightarrow n) \right] \cr
\noalign{\kern 5pt}
&&\hskip -1.cm
=-g_w\sqrt{\frac{kL}{z_L^2-1}} 
\frac{1}{\sqrt{r_{W^\m} r_{W^\n}}}\sin\theta_H\cos\theta_H  \cr
\noalign{\kern 5pt}
&&\hskip -0.cm
\times \int_1^{z_L} dz 
\Big\{ \big(\partial_z \hat{S}_{W^\m}\big)C_{W^\n}
-\big(\partial_z C_{W^\m}\big)\hat{S}_{W^\n} 
+\big( m\longleftrightarrow n\big) \Big\} ~.
\label{HWW1}
\eeqn
Numerical values of  $g_{HW^\m W^\n}$ are given in Table \ref{tbl:HWW}.

\begin{table}[thbp]\centering \footnotesize
\caption{
$g_{HW^\m W^\n}/g_w\cos\theta_H$ in the unit of GeV written by three significant figures. The values smaller than $O(10)$ are abbreviated.}\label{tbl:HWW}
\begin{tabular}{|c|cccccccc|}\hline
   &0&1&2&3&4&5&6&7\\ \hline
 0 &  80.0 & 2.55$\times 10^2$ & $O(1)$ & 45.4 & $O(10^{-1})$ & 20.7 & $O(10^{-1})$ & 10.4 \\ 
 1 &  2.55$\times 10^2$ & -3.50$\times 10^2$ &  1.39$\times 10^4$ & -1.96$\times 10^2$
   &  1.40$\times 10^3$ & $O(1)$             &  2.28$\times 10^2$ & -24.1   \\ 
 2 &  $O(1)$            &  1.39$\times 10^4$ &  5.62$\times 10^2$ &  2.06$\times 10^4$
   &  2.87$\times 10^2$ &  3.04$\times 10^3$ &  $O(1)$            &  1.66$\times 10^3$  \\ 
 3 &  45.4              & -1.96$\times 10^2$ &  2.06$\times 10^4$ & -8.40$\times 10^2$
   &  2.94$\times 10^4$ & -4.17$\times 10^2$ &  3.54$\times 10^3$ & $O(1)$              \\ 
 4 &  $O(10^{-1})$      &  1.40$\times 10^3$ &  2.87$\times 10^2$ &  2.93$\times 10^4$
   &  1.07$\times 10^3$ &  3.49$\times 10^4$ &  5.11$\times 10^2$ &  4.51$\times 10^3$  \\ 
 5 &  20.7              & $O(1)$             &  3.04$\times 10^3$ & -4.17$\times 10^2$
   &  3.49$\times 10^4$ & -1.36$\times 10^3$ &  4.46$\times 10^4$ & -6.40$\times 10^2$  \\ 
 6 &  $O(10^{-1})$      &  2.28$\times 10^2$ &  $O(1)$            &  3.54$\times 10^3$ 
   &  5.11$\times 10^2$ &  4.46$\times 10^4$ &  1.60$\times 10^3$ &  4.88$\times 10^4$  \\ 
 7 &  10.4              & -24.1              &  1.66$\times 10^3$ & $O(1)$
   &  4.51$\times 10^3$ & -6.40$\times 10^2$ &  4.88$\times 10^4$ & -1.90$\times 10^3$  \\ 
\hline
\end{tabular}\end{table}

\subsection{$ZW^\m W_R^\n$ coupling}
The $ZW^\m W_R^\n$ coupling in
\beqn
&&i\sum_{m,n}g_{ZW^\m W_R^\n}\eta^{\mu \rho}\eta^{\nu\sigma}
\Big\{(\partial_\mu Z_\nu -\partial_\nu Z_\mu) (W^{+(m)}_\rho W^{-(n)}_{R\:\sigma}+W^{-(m)}_\rho W^{+(n)}_{R\:\sigma}) \nonumber\\
&&\qquad
+(\partial_\mu W^{-(m)}_\nu -\partial_\nu W^{-(m)}_\mu) Z_\rho W^{+(n)}_{R\:\sigma}
-(\partial_\mu W^{+(m)}_\nu -\partial_\nu W^{+(m)}_\mu) Z_\rho W^{-(n)}_{R\:\sigma} \cr
&&\qquad
+(\partial_\mu W^{-(n)}_{R\:\nu} -\partial_\nu W^{-(n)}_{R\:\mu}) Z_\rho W^{+(m)}_{\sigma} 
-(\partial_\mu W^{+(n)}_{R\:\nu} -\partial_\nu W^{+(n)}_{R\:\mu}) Z_\rho W^{-(m)}_{\sigma}
\Big\}
\eeqn
is given by
\beqn
&&\hskip -1.cm
g_{ZW^\m W^\n_R} = g_w \sqrt{L} \int_1^{z_L} \frac{dz}{kz}
\biggl\{ h_Z^L h_{W^\m}^Lh_{W^\n_R}^L+h_Z^R h_{W^\m}^Rh_{W^\n_R}^R
+\hat{h}_Z\hat{h}_{W^\m}\frac{h_{W^\n_R}^L+h_{W^\n_R}^R}{2}\bigg\}  \cr
\noalign{\kern 5pt}
&&\hskip .5cm
=g_w \cos\theta_W \frac{\sqrt{L}}{\sqrt{r_Z\:r_{W^\m}\:r_{W^\n_R}}} 
\int_1^{z_L} \frac{dz}{kz}\frac{1}{\sqrt{2}}  \cr
\noalign{\kern 5pt}
&&\hskip 1.cm 
\times \frac{\sin^2\theta_H}{\cos^2\theta_W}
\frac{\cos\theta_H}{\sqrt{1+\cos^2\theta_H}}
\biggl\{C_ZC_{W^\m}C_{W^\n_R}-\hat{S}_Z\hat{S}_{W^\m}C_{W^\n_R}\bigg\} ~.
\label{ZWWR1}
\eeqn
Numerical values of  $g_{ZW^\m W_R^\n}$ are given in Table \ref{tbl:ZWWR}.

\begin{table}[thbp]\centering \footnotesize
\caption{$g_{ZW^\m W^\n_R}/g_w\cos\theta_W$. 
Only the values larger than $O(10^{-3})$ are shown and written by two significant figures.}\label{tbl:ZWWR}
\begin{tabular}{|cc|cccccccc|}\hline
	&&&&&$m$&&&&\\
	&&0&1&2&3&4&5&6&7 \\ \hline
    &1& $O(10^{-4})$ &   \ 0.004    &   -0.027     & $O(10^{-5})$ & $O(10^{-4})$   
      & $O(10^{-6})$ & $O(10^{-5})$ & $O(10^{-6})$ \\
$n$ &2& $O(10^{-5})$ & $O(10^{-5})$ &  \ 0.025     &  \ 0.004     &   -0.027  
      & $O(10^{-5})$ & $O(10^{-4})$ & $O(10^{-6})$ \\
    &3& $O(10^{-6})$ & $O(10^{-6})$ &  \ 0.001     & $O(10^{-4})$ &  \ 0.026 
      &  \ 0.004     &   -0.027     & $O(10^{-5})$ \\
    &4& $O(10^{-6})$ & $O(10^{-6})$ & $O(10^{-4})$ & $O(10^{-6})$ &  \ 0.001
      & $O(10^{-4})$ &  \ 0.027     &  \ 0.004     \\
\hline\end{tabular}
\end{table}

\subsection{$\gamma ZW^\m W_R^\n$ coupling}
Similarly $\gamma ZW^\m W_R^\n$ coupling is contained in 
\beqn
&&\hskip -1.cm
(g_A)^2\int_1^{z_L}\frac{dz}{kz}\eta^{\mu \rho}\eta^{\nu\sigma}\cr
&&\times\text{Tr}\Big[[\hat{A}^{\gamma(A)}_\mu, \hat{W}^+_\nu]
\left([\hat{Z}_\rho, \hat{W}^-_{R\:\sigma}]-[\hat{Z}_\sigma, \hat{W}^-_{R\:\rho}]\right)+[\hat{A}^{\gamma(A)}_\mu, \hat{W}^-_\nu]
\left([\hat{Z}_\rho, \hat{W}^+_{R\:\sigma}]-[\hat{Z}_\sigma, \hat{W}^+_{R\:\rho}]\right) \cr
&&\hskip 1.cm
[\hat{A}^{\gamma(A)}_\mu, \hat{W}^-_{R\:\nu}]
\left([\hat{Z}_\rho, \hat{W}^+_{R\:\sigma}]-[\hat{Z}_\sigma, \hat{W}^+_\rho]\right)+[\hat{A}^{\gamma(A)}_\mu, \hat{W}^-_{R\:\nu}]
\left([\hat{Z}_\rho, \hat{W}^+_{R\:\sigma}]-[\hat{Z}_\sigma, \hat{W}^+_\rho]\right)\Big]\cr
\noalign{\kern 5pt}
&& \hskip -1.cm
\supset \sum_{m,n}g_{\gamma ZW^\m W_R^\n}\eta^{\mu \rho}\eta^{\nu\sigma} \nonumber\\
&&\times\big\{
 A^{\gamma}_\mu W^{+(m)}_\nu (Z_\rho W^{-(n)}_{R\:\sigma} - Z_\sigma W^{-(n)}_{R\:\rho})
+A^{\gamma}_\mu W^{-(n)}_\nu (Z_\rho W^{+(m)}_{R\:\sigma} - Z_\sigma W^{+(m)}_{R\:\rho})\cr 
&& \hskip 0.3cm
+A^{\gamma}_\mu W^{+(m)}_{R\:\nu} (Z_\rho W^{-(n)}_{\sigma} - Z_\sigma W^{-(n)}_{\rho})
+A^{\gamma}_\mu W^{-(n)}_{R\:\nu} (Z_\rho W^{+(m)}_{\sigma} - Z_\sigma W^{+(m)}_{\rho})\big\} 
\eeqn
so that
\begin{align}
g_{\gamma ZW^\m W^\n_R} 
&= -g_w^2 \sqrt{L} \int_1^{z_L} \frac{dz}{kz}
\biggl\{ h_{\gamma^{(0)}}^Lh_{W^\m}^Lh_Z^Lh_{W^\n_R}^L+h_{\gamma^{(0)}}^Rh_{W^\m}^Rh_Z^R h_{W^\n_R}^R\nonumber\\
&\hskip 3.cm
+\frac{1}{2}\left(h_{\gamma^{(0)}}^L+h_{\gamma^{(0)}}^R\right)\hat{h}_{W^\m}
\frac{1}{2}\hat{h}_Z\left(h_{W^\n_R}^L+h_{W^\n_R}^R\right)\bigg\}  \nonumber\\
&=-eg_w g_{ZW^\m W^\n_R}.
\end{align}
The relation $g_{\gamma ZW^\m W_R^\n}= -e g_{ZW^\m W_R^\n}$ follows from the gauge invariance as well.

\subsection{$HW^\m W^\n_R$ and $HW^\m_R W^\n_R$ coupling}
Similarly the $HW^\m W^\n_R$ coupling contained in
\begin{align}
& -i g_A k^2 \int_1^{z_L}\frac{dz}{kz} 
\text{Tr}\left[\partial_z \hat{W}^-_{R\:\mu} \big[\hat{W}^+_\nu, \hat{H}\big]+
\partial_z \hat{W}^+_\mu \big[\hat{W}^-_{R\:\nu}, \hat{H}\big]\right]\eta^{\mu\nu}\nonumber\\
&\supset -\sum_{m,n}g_{HW^\m W^\n_R} H W^{+(m)}_\mu W^{-(n)}_{R\:\nu}\; \eta^{\mu\nu}
\end{align}
is given by 
\beqn
&&\hskip -1.cm
g_{HW^\m W^\n_R}
=i g_A k^2 \int_1^{z_L}\frac{dz}{kz} \cr
\noalign{\kern 5pt}
&&\hskip .5cm
\times \frac{i}{2}u_H(z)
\left[\partial_z\big(h^L_{W^\n_R}-h^R_{W^\n_R}\big)\hat{h}_{W^\m}
-\big(\partial_z \hat{h}_{W^\m}\big)\big(h^L_{W^\n_R}-h^R_{W^\n_R}\big)\right] \cr
\noalign{\kern 5pt}
&&\hskip -.5cm
=-g_w\sqrt{\frac{kL}{z_L^2-1}} 
\frac{1}{\sqrt{r_{W^\m} r_{W^\n_R}}}(-\sin\theta_H) \cr
\noalign{\kern 5pt}
&&\hskip .5cm
\times \int_1^{z_L} dz 
\left[\big(\partial_zC_{W^\n_R}\big)\hat{S}_{W^\m}
-\big(\partial_z \hat{S}_{W^\m}\big)C_{W^\n_R}\right] ~. 
\label{HWWR1}
\eeqn
Numerical values of  $g_{HW^\m W_R^\n}$ are given in Table \ref{tbl:HWWR}. 
The $HW_R^\m W_R^\n$ couplings vanish for all $m, n$ 
as a result of the Lie algebra.

\begin{table}[thbp]\centering \footnotesize
\caption{$g_{HW^\m W^\n_R}/g_w$ in the unit of GeV written by three significant figures. 
The values smaller than $O(10)$ are abbreviated.}\label{tbl:HWWR}
\begin{tabular}{|cc|cccccccc|}\hline
	&&&&&$m$&&&&\\
	&&0&1&2&3&4&5&6&7 \\ \hline
     &1&  266  & -168  &  1.27$\times 10^4$ & -69.0  &  974 & $O(1)$ &  280 & $O(1)$ \\
 $n$ &2&  50.5 & -123  &  2.13$\times 10^4$ & -411   &  2.76$\times 10^4$ & -166 &  2.63$\times 10^3$ & $O(1)$\\
     &3&  20.6 & -10.6 &  3.60$\times 10^3$ & -247   &  3.62$\times 10^4$ & -670 &  4.22$\times 10^4$ & -264 \\
     &4&  11.1 & -16.0 &  1.56$\times 10^3$ & -13.2  &  5.49$\times 10^3$ & -371 &  5.08$\times 10^4$ & -940 \\
\hline\end{tabular}
\end{table}

\subsection{$Zt^\m t^\n$ coupling}
The $Z$ couplings of the top quark tower are found from
\begin{align}
& \sum_{a=1}^2 
\int_1^{z_L} dz\sqrt{G}
\longbar{\Psi}_a\left(-ig_A A_\mu-ig_BQ_{X_a} B_\mu\right) z\gamma^\mu \Psi_a \nonumber\\
&\supset-ig_w Z_\mu\sqrt{L} 
\int_1^{z_L} \frac{dz}{k}\nonumber\\&\times
\bigg\{\frac{h^L_Z}{2}
\left(\bar{\tilde{U}}\gamma^\mu \tilde{U}-\bar{\tilde{B}}\gamma^\mu \tilde{B}+\bar{\tilde{t}}\gamma^\mu \tilde{t}\right)
+\frac{h^R_Z}{2}
\left(\bar{\tilde{U}}\gamma^\mu \tilde{U}+\bar{\tilde{B}}\gamma^\mu \tilde{B}-\bar{\tilde{t}}\gamma^\mu \tilde{t}\right)
\nonumber\\&\qquad
+\frac{\hat{h}_Z}{2}
\left(\bar{\tilde{B}}\gamma^\mu \tilde{t'}+\bar{\tilde{t}}\gamma^\mu \tilde{t'}
+\bar{\tilde{t'}}\gamma^\mu \tilde{B}+\bar{\tilde{t'}}\gamma^\mu \tilde{t}\right)
\nonumber\\&\qquad
+Q_{X_1}t_\phi h_Z^B\left(\bar{\tilde{B}}\gamma^\mu \tilde{B}
+\bar{\tilde{t}}\gamma^\mu \tilde{t}+\bar{\tilde{t'}}\gamma^\mu \tilde{t'}\right)
+Q_{X_2}t_\phi h_Z^B\bar{\tilde{U}}\gamma^\mu \tilde{U}
\bigg\} ~.
\end{align}
The $Zt^\m t^\n$ couplings are found to be
\begin{align}
& -i\frac{g_w}{\cos\theta_W}
\sum_{m,n}Z_\mu(-i)\; 
\bar{t}^\m_L\gamma^\mu t^\n_L\frac{\sqrt{L}}{\sqrt{2 r_Z}} 
\int_1^{z_L} dz\nonumber\\&
\times\bigg\{ C_Zf^\m_{U_L}f^\n_{U_L}+\cos\theta_H C_Z
\left(-f^\m_{B_L}f^\n_{B_L}+f^\m_{t_L}f^\n_{t_L}\right)
\nonumber\\
&\qquad+\frac{-\sqrt{2}\sin\theta_H}{2}\hat{S}_Z
\left(f^\m_{B_L}f^\n_{t'_L}+f^\m_{t_L}f^\n_{t'_L}+f^\m_{t'_L}f^\n_{B_L}+f^\m_{t'_L}f^\n_{t_L}\right)
\nonumber\\
&\qquad
-2\sin^2\theta_WC_Z Q_{X_1} 
\left(f^\m_{U_L}f^\n_{U_L}+f^\m_{B_L}f^\n_{B_L}+f^\m_{t_L}f^\n_{t_L}+f^\m_{t'_L}f^\n_{t'_L}
\right)\bigg\} 
\end{align}
for the left-handed component $t^\n_L$ and a similar expression for $t^\n_R$.
Noting that $Q_{X_1}  = \frac{2}{3}$ and $Q_{X_2}  = - \onethird$, 
one finds that
\beeq
Z_\mu \sum_{m,n}
\Big\{ g_{Zt^\m_L t^\n_L}  \bar{t}^\m_L \gamma^\mu t^\n_L +
g_{Zt^\m_R t^\n_R}  \bar{t}^\m_R  \gamma^\mu t^\n_R \Big\} 
\eneq
where 
\beqn
&&\hskip -1.cm
g_{Zt^\m_L t^\n_L}=\frac{g_w}{\cos\theta_W}
\frac{\sqrt{2L}}{\sqrt{r_Zr_{t^\m} r_{t^\n}}}   \int_1^{z_L} dz \cr
\noalign{\kern 5pt}
&&\hskip -1.cm
\times \bigg[ \Big(\frac{\tilde{\mu}^2}{\mu_2^2}+c^2_H\Big) C_L^\m C_L^\n C_Z
+\frac{s^2_H}{2} \bigg( C_L^\m \frac{C_L^\n(1)}{S_L^\n(1)} S_L^\n+
\frac{C_L^\m(1)}{S_L^\m(1)} S_L^\m C_L^\m \bigg) \hat{S}_Z \cr
\noalign{\kern 5pt}
&&\hskip -1.cm
- \twothird \sin^2\theta_W
 \bigg\{\bigg(2\frac{\tilde{\mu}^2}{{\mu_2}^2}+c^2_H+1\bigg)C_L^\m C_L^\n 
+s^2_H \frac{C_L^\m(1)}{S_L^\m(1)} S_L^\m \frac{C_L^\n(1)}{S_L^\n(1)} S_L^\n
\bigg\}C_Z\bigg] , \cr
\noalign{\kern 10pt} 
&&\hskip -1.cm
g_{Zt^\m_R t^\n_R}= \frac{g_w}{\cos\theta_W}
\frac{\sqrt{2}\sqrt{L}}{\sqrt{r_Zr_{t^\m} r_{t^\n}}} \int_1^{z_L} dz \cr
\noalign{\kern 5pt}
&&\hskip -1.cm
\times \bigg[ \Big(\frac{\tilde{\mu}^2}{\mu_2^2}+c^2_H\Big)S_R^\m S_R^\n C_Z
+\frac{s^2_H}{2}\bigg( S_R^\m \frac{C_L^\n(1)}{S_L^\n(1)} C_R^\n+
\frac{C_L^\m(1)}{S_L^\m(1)} C_R^\m S_R^\m \bigg) \hat{S}_Z \cr
\noalign{\kern 5pt}
&&\hskip -1.cm
- \twothird \sin^2\theta_W
\bigg\{\bigg(2\frac{\tilde{\mu}^2}{{\mu_2}^2}+c^2_H+1\bigg)S_R^\m S_R^\n 
+s^2_H \frac{C_L^\m(1)}{S_L^\m(1)} C_R^\m \frac{C_L^\n(1)}{S_L^\n(1)} C_R^\n
\bigg\} C_Z \bigg] .
\label{Ztt1}
\eeqn
Therefore the vector and axial vector coupling are written by
\begin{equation}
g^V_{Zt^\m t^\n}= \frac{g_{Zt^\m_L t^\n_L} + g_{Zt^\m_R t^\n_R}}{2}, \qquad 
g^A_{Zt^\m t^\n}= \frac{g_{Zt^\m_L t^\n_L} - g_{Zt^\m_R t^\n_R}}{2}
\end{equation}
Numerical values of  $g^V_{Zt^\m t^\n}$ are given in Table \ref{tbl:Ztt}.

\begin{table}[thbp]\centering \footnotesize
\caption{$g^V_{Zt^\m t^\n} = \onehalf \{ g_{Zt^\m_L t^\n_L} + g_{Zt^\m_R t^\n_R} \}$
in the unit of $g/\cos\theta_W$. The values larger than $O(10^{-3})$ are shown. 
$g^A_{Zt^\m t^\n} = \onehalf \{ g_{Zt^\m_L t^\n_L} - g_{Zt^\m_R t^\n_R} \}$ in the unit of 
$g/\cos\theta_W$ is smaller than $O(10^{-3})$ in the range of $m,n \leq 10$, 
except for $g^A_{Zt^{(0)} t^{(0)}}=-0.2501$.}\label{tbl:Ztt}
\begin{tabular}{|c|cccccccc|}\hline
   &0&1&2&3&4&5&6&7\\ \hline
 0 &    0.095     &    -0.008    &    0.001     & $O(10^{-4})$ & $O(10^{-5})$ & $O(10^{-4})$
   & $O(10^{-5})$ & $O(10^{-5})$ \\
 1 &   -0.008     &    0.337     &    0.059     & $O(10^{-4})$ &    0.002     & $O(10^{-5})$
   & $O(10^{-6})$ & $O(10^{-6})$ \\
 2 &    0.001     &    0.059     &   -0.149     &   -0.010     & $O(10^{-5})$ & $O(10^{-4})$
   & $O(10^{-5})$ & $O(10^{-4})$ \\
 3 & $O(10^{-4})$ & $O(10^{-4})$ &   -0.010     &    0.338     &    0.056     & $O(10^{-5})$
   &    0.002     & $O(10^{-6})$ \\
 4 & $O(10^{-5})$ &    0.002     & $O(10^{-5})$ &    0.056     &   -0.149     &  -0.010
   & $O(10^{-5})$ & $O(10^{-4})$ \\
 5 & $O(10^{-4})$ & $O(10^{-5})$ & $O(10^{-4})$ & $O(10^{-5})$ &   -0.010     &   0.338
   &    0.056     & $O(10^{-4})$ \\
 6 & $O(10^{-5})$ & $O(10^{-6})$ & $O(10^{-5})$ &    0.002     & $O(10^{-5})$ &   0.056 
   &   -0.150     &   -0.010     \\
 7 & $O(10^{-5})$ & $O(10^{-6})$ & $O(10^{-4})$ & $O(10^{-6})$ & $O(10^{-4})$ & $O(10^{-4})$
   &   -0.010     &    0.338     \\
\hline\end{tabular}\end{table}

\subsection{$Ht^\m t^\n$ coupling}
The Higgs couplings of the top quark tower are contained in
\begin{align}
& 
\int_1^{z_L} dz\sqrt{G}
\longbar{\Psi}_1\left(-ig_A kzA_z \right)\gamma^5 \Psi_1\nonumber\\
&\supset -ig_w\sqrt{L}H \int_1^{z_L} dz \; u_H(z)
\frac{i}{2}\left(\bar{B}\gamma^5t'-\bar{t}\gamma^5t'-\bar{t'}\gamma^5B+\bar{t'}\gamma^5t\right)
\nonumber\\
&=\frac{g_wk\sqrt{L}}{2}
\sum_{m,n} H it_L^{(m)\dagger} t^\n_R\nonumber\\
&\times \int_1^\infty{z_L} dz \; u_H(z)
\left(f^\m_{B_L}f^\n_{t'_R}-f^\m_{t_L}f^\n_{t'_R}-f^\m_{t'_L}f^\n_{B_R}+f^\m_{t'_L}f^\n_{t_R}\right)
+(L\leftrightarrow R)\nonumber\\
&= i \sum_{m,n} H \, 
\left( g_{Ht_L^{(m)\dagger} t^\n_R}   t_L^{(m)\dagger} t^\n_R 
-   g_{Ht_R^{(m)\dagger} t^\n_L}   t_R^{(m)\dagger} t^\n_L \right) ~.
\end{align}
One finds that 
\beqn
&&\hskip -1.cm
g_{Ht_R^{(m)\dagger} t^\n_L} = g_{Ht_L^{(n)\dagger} t^\m_R} \cr
\noalign{\kern 10pt}
&&\hskip -1.cm
= -\frac{g_wk\sqrt{L}}{2} \int_1^{z_L} dz \; u_H(z)
\left(f^\m_{B_R}f^\n_{t'_L}-f^\m_{t_R}f^\n_{t'_L}
-f^\m_{t'_R}f^\n_{B_L}+f^\m_{t'_R}f^\n_{t_L} \right) \cr
\noalign{\kern 10pt}
&&\hskip -1.cm
= -\frac{g_w\sqrt{kL}}{2}
\frac{\sqrt{2}}{\sqrt{(z_L^2-1) r_{t^\m} r_{t^\n}}} \cr
\noalign{\kern 10pt}
&&\hskip -.5cm
\times \int_1^{z_L} dz \; z
\bigg\{ -\sqrt{2}c_H S^\m_R s_H\frac{C_L^\n(1)}{S_L^\n(1)}S_L^\n
+\sqrt{2}s_H \frac{C_L^\m(1)}{S_L^\m(1)}C^\m_R c_H C_L^\n\bigg\} .
\label{Htt2}
\eeqn
We denote 
\beeq
y_{t^\m t^\n} =\frac{g_{Ht_L^{(m)\dagger}t^\n_R}+g_{Ht_R^{(m)\dagger}t^\n_L}}{2}  ~~,~~
\hat{y}_{t^\m t^\n}=\frac{-g_{Ht_L^{(m)\dagger}t^\n_R}+g_{Ht_R^{(m)\dagger}t^\n_L}}{2} ~~.
\label{Htt3}
\eneq
For $m=n$ 
\beqn
&&\hskip -1.cm
y_{t^\m t^\m} = -\frac{g_wk\sqrt{kL}}{2}
\frac{\sqrt{z_L^2-1}}{r_{t^\n}}s_H c_H \frac{C_L^\m(1)}{S_L^\m(1)} ~, \cr
\noalign{\kern 10pt}
&&\hskip -1.cm
\hat{y}_{t^\m t^\m} = 0 ~.
\eeqn
Numerical values of  $y_{t^\m t^\n}$ and $\hat{y}_{t^\m t^\n}$ are given in Table \ref{tbl:ytt} and \ref{tbl:yhtt}, respectively.

\begin{table}[thbp]\centering \footnotesize
\caption{$y_{t^\m t^\n}$ in the unit of $y_t\cos\theta_H$. Only the values larger than $O(10^{-3})$ are shown and written by three significant figures.}\label{tbl:ytt}
\begin{tabular}{|c|cccccccc|}\hline
&0&1&2&3&4&5&6&7\\ \hline
 0 &  1.00  &     0.517    &     0.188    &     0.049    &    -0.010    &     0.044    &     0.025    &     0.013    \\
 1 &  0.517 &    -0.225    &     1.04     &    -0.090    &     0.234    & $O(10^{-4})$ & $O(10^{-4})$ &    -0.010    \\
 2 &  0.188 &     1.04     &     0.226    &     0.674    &     0.088    &     0.034    & $O(10^{-4})$ &     0.057    \\
 3 &  0.049 &    -0.090    &     0.694    &    -0.217    &     1.05     &    -0.087    &     0.244    & $O(10^{-4})$ \\
 4 & -0.010 &     0.234    &     0.088    &     1.05     &     0.217    &     0.670    &     0.087    &     0.028    \\
 5 &  0.044 & $O(10^{-4})$ &     0.034    &    -0.087    &     0.670    &    -0.214    &     1.05     &    -0.087    \\
 6 &  0.025 & $O(10^{-4})$ & $O(10^{-4})$ &     0.244    &     0.087    &     1.05     &     0.215    &     0.667    \\
 7 &  0.013 &    -0.010    &     0.057    & $O(10^{-4})$ &     0.028    &    -0.087    &     0.667    &    -0.213    \\
\hline\end{tabular}\end{table}

\begin{table}[thbp]\centering \footnotesize
\caption{$\hat{y}_{t^\m t^\n}$ in the unit of $y_t\cos\theta_H$. Only the values larger than $O(10^{-3})$ are shown and written by three significant figures.}\label{tbl:yhtt}
\begin{tabular}{|c|cccccccc|}\hline
   &0&1&2&3&4&5&6&7\\ \hline
 0 &    0   &    -0.529    &    0.091     &    -0.043    &    -0.015    &    -0.049    &     0.015    &    -0.011    \\
 1 &  0.529 &       0      &   -0.040     &     0.014    &    -0.005    & $O(10^{-4})$ & $O(10^{-5})$ &     0.002    \\
 2 & -0.091 &     0.040    &      0       &    -0.119    &    -0.012    &    -0.011    & $O(10^{-4})$ &    -0.026    \\
 3 &  0.043 &     0.012    &    0.119     &       0      &    -0.024    &     0.008    &    -0.014    & $O(10^{-4})$ \\
 4 &  0.015 & $O(10^{-3})$ &    0.012     &     0.024    &       0      &    -0.060    &    -0.007    &    -0.005    \\
 5 &  0.049 & $O(10^{-4})$ &    0.011     &    -0.008    &     0.062    &       0      &    -0.017    &     0.006    \\
 6 & -0.015 & $O(10^{-4})$ & $O(10^{-4})$ &     0.014    &     0.007    &     0.017    &       0      &    -0.040    \\
 7 &  0.011 & $O(10^{-3})$ &    0.026     & $O(10^{-4})$ &     0.005    &     0.006    &    0.040     &       0      \\
\hline\end{tabular}\end{table}

\subsection{$ZF^\m F^\n$ coupling}
The $Z$ couplings of the dark fermion tower are given by
\beqn
&&\hskip -1.cm
\int_1^{z_L} dz\sqrt{G}
\longbar{\Psi}_F\left(-ig_AA_\mu-ig_BQ_{X_F}B_\mu\right)z\gamma^\mu \Psi_F \cr
\noalign{\kern 10pt}
&&\hskip -1.cm
=-iZ_\mu\frac{g_w}{\cos\theta_W} \frac{\sqrt{L}}{\sqrt{2}\sqrt{r_Z}}\int_1^{z_L} dz \sum_{m,n} \cr
\noalign{\kern 10pt}
&&\hskip -.5cm
\times \Big[\Big\{\left(1+\cos\theta_H\right)C(z) f^{*(m)}_{l L}f^\n_{lL} 
+\left(1-\cos\theta_H\right)C(z)f^{*(m)}_{rL}f^\n_{rL} \cr
\noalign{\kern 5pt}
&&\hskip .5cm
-\sin\theta_H\hat{S}(z)\left(if^{*(m)}_{lL}f^\n_{rL}-if^{*(m)}_{rL}f^\n_{lL}\right)\Big\}
\Fbar^{(m)}_L\gamma^\mu I_3 F^{(n)}_L \cr
\noalign{\kern 5pt}
&&\hskip .5cm
-2\sin^2\theta_W\left(f^{*(m)}_{lL}f^\n_{lL}+f^{*(m)}_{rL}f^\n_{rL}\right)
\Fbar^{(m)}_L\gamma^\mu (I_3+Q_{X_F})F^{(n)}_L +(L\rightarrow R) \Big] \cr
\noalign{\kern 10pt}
&&\hskip -1.cm
= -iZ_\mu \sum_{m,n}
\Big\{ g_{ZF^{+(m)}_LF^{+(n)}_L}\Fbar^{+(m)}_L\gamma^\mu F^{+(n)}_L
+g_{ZF^{+(m)}_RF^{+(n)}_R}\Fbar^{+(m)}_R\gamma^\mu F^{+(n)}_R \cr
\noalign{\kern 5pt}
&&\hskip 1.5cm
+g_{ZF^{0(m)}_LF^{0(n)}_L}\Fbar^{0(m)}_L\gamma^\mu F^{0(n)}_L
+g_{ZF^{0(m)}_RF^{0(n)}_R}\Fbar^{0(m)}_R\gamma^\mu F^{0(n)}_R \Big\} ~,
\label{ZFF1}
\eeqn
where $F^\n_L$ is a abbreviation of the doublet $(F^{+(n)}_L, F^{0(n)}_L)^T$ 
and $I_3$ is a isospin operator. 
For $Q_{X_F}=1/2$ one finds that
\beqn
&&\hskip -1.cm
g_{ZF^{+(m)}_LF^{+(n)}_L} =\frac{g_w}{\cos\theta_W}\frac{\sqrt{L}}{2\sqrt{2}} \frac{1}{\sqrt{r_Z}}
\int_1^{z_L} dz \frac{1}{\sqrt{r_{F^\m}r_{F^\n}}} \cr
\noalign{\kern 10pt}
&&\hskip -1.cm
\times \bigg\{\big(1+\cos\theta_H-4\sin^2\theta_W\big)C_Z(z) \sin^2\frac{\theta_H}{2}
S^\m_L(1)C^\m_L(z) S^\n_L(1)C^\n_L(z) \cr
\noalign{\kern 10pt}
&&\hskip -.7cm
+\big(1-\cos\theta_H-4\sin^2\theta_W\big)C_Z(z) \cos^2\frac{\theta_H}{2}
C^\m_L(1)S^\m_L(z) C^\n_L(1)S^\n_L(z) \cr
\noalign{\kern 10pt}
&&\hskip -.7cm
-\sin\theta_H\hat{S}(z)\sin\frac{\theta_H}{2}\cos\frac{\theta_H}{2}  \cr
\noalign{\kern 10pt}
&&\hskip -.0cm
\times\left(S^\m_L(1)C^\m_L(z)C^\n_L(1)S^\n_L(z)+C^\m_L(1)S^\m_L(z)S^\n_L(1)C^\n_L(z)\right) \bigg\} , \cr
\noalign{\kern 10pt}
&&\hskip -1.cm
g_{ZF^{0(m)}_LF^{0(n)}_L} =-\frac{g_w}{\cos\theta_W}\frac{\sqrt{L}}{2\sqrt{2}} \frac{1}{\sqrt{r_Z}}
\int_1^{z_L} dz \frac{1}{\sqrt{r_{F^\m}r_{F^\n}}} \cr
\noalign{\kern 10pt}
&&\hskip -1.cm
\times \bigg\{ \big(1+\cos\theta_H\big)C_Z(z) \sin^2\frac{\theta_H}{2}
S^\m_L(1)C^\m_L(z) S^\n_L(1)C^\n_L(z) \cr
\noalign{\kern 10pt}
&&\hskip -.7cm
+\big(1-\cos\theta_H\big)C_Z(z) \cos^2\frac{\theta_H}{2}C^\m_L(1)S^\m_L(z) C^\n_L(1)S^\n_L(z) \cr
\noalign{\kern 10pt}
&&\hskip -.7cm
-\sin\theta_H\hat{S}(z)\sin\frac{\theta_H}{2}\cos\frac{\theta_H}{2}  \cr
\noalign{\kern 10pt}
&&\hskip -.0cm
\times\left(S^\m_L(1)C^\m_L(z)C^\n_L(1)S^\n_L(z)+C^\m_L(1)S^\m_L(z)S^\n_L(1)C^\n_L(z)\right) \bigg\}.
\label{ZFF2}
\eeqn
$g_{ZF^{+(m)}_RF^{+(n)}_R}$ and $g_{ZF^{0(m)}_RF^{0(n)}_R}$ 
are obtained from the formulas for $g_{ZF^{+(m)}_LF^{+(n)}_L}$ and $g_{ZF^{0(m)}_LF^{0(n)}_L}$
by replacing $C_L(z)$ to $S_R(z)$ and $S_L(z)$ to $C_R(z)$,  respectively.
Numerical values of  $g^V_{ZF^{+(m)}F^{+(n)}}$ and $g^V_{ZF^{0(m)}F^{0(n)}}$ are given in Table \ref{tbl:ZFF+} and \ref{tbl:ZFF0}, respectively.

\begin{table}[thbp]\centering \footnotesize
\caption{
$g^V_{ZF^{+(m)} F^{+(n)}} = \onehalf \{ g_{ZF^{+(m)}_LF^{+(n)}_L} + g_{ZF^{+(m)}_RF^{+(n)}_R} \}$ 
in the unit of $g/\cos\theta_W$. Only the value larger than $O(10^{-3})$ are shown and written by three significant figures.}\label{tbl:ZFF+}
\begin{tabular}{|c|ccccccc|}\hline
   &1&2&3&4&5&6&7\\ \hline
 1 &   -0.230     &    0.021     & $O(10^{-5})$ &   -0.001     & $O(10^{-6})$   & $O(10^{-4})$ & $O(10^{-6})$ \\
 2 &    0.021     &    0.267     &    0.009     & $O(10^{-6})$ & $O(10^{-5})$   & $O(10^{-6})$ & $O(10^{-5})$ \\
 3 & $O(10^{-5})$ &    0.009     &   -0.230     &    0.024     & $O(10^{-6})$   &   -0.001     & $O(10^{-6})$ \\
 4 &   -0.001     & $O(10^{-6})$ &    0.024     &    0.267     &    0.009       & $O(10^{-6})$ & $O(10^{-4})$ \\
 5 & $O(10^{-6})$ & $O(10^{-5})$ & $O(10^{-6})$ &    0.009     &   -0.229       &    0.025     & $O(10^{-6})$ \\
 6 & $O(10^{-4})$ & $O(10^{-6})$ &   -0.001     & $O(10^{-6})$ &    0.025       &    0.267     &    0.009     \\
 7 & $O(10^{-6})$ & $O(10^{-5})$ & $O(10^{-6})$ & $O(10^{-4})$ & $O(10^{-6})$   &    0.009     &   -0.229     \\
\hline
\end{tabular}\end{table}
\begin{table}[thbp]\centering \footnotesize
\caption{
$g^V_{ZF^{0(m)} F^{0(n)}}= \onehalf \{  g_{ZF^{0(m)}_LF^{0(n)}_L} + g_{ZF^{0(m)}_RF^{0(n)}_R} \}$
in the unit of $g/\cos\theta_W$. Only the value larger than $O(10^{-3})$ are shown and written by three significant figures.}\label{tbl:ZFF0}
\begin{tabular}{|c|ccccccc|}\hline
   &1&2&3&4&5&6&7\\ \hline
 1 &   -0.002     &   -0.021     & $O(10^{-5})$ &    0.001     & $O(10^{-6})$   & $O(10^{-4})$ & $O(10^{-6})$ \\
 2 &   -0.021     &   -0.498     &   -0.009     & $O(10^{-5})$ & $O(10^{-5})$   & $O(10^{-6})$ & $O(10^{-5})$ \\
 3 & $O(10^{-5})$ &   -0.009     &   -0.002     &   -0.024     & $O(10^{-5})$   &    0.001     & $O(10^{-6})$ \\
 4 &    0.001     & $O(10^{-5})$ &   -0.024     &   -0.498     &   -0.009       & $O(10^{-5})$ & $O(10^{-4})$ \\
 5 & $O(10^{-6})$ & $O(10^{-5})$ & $O(10^{-5})$ &   -0.009     &   -0.002       &   -0.025     & $O(10^{-5})$ \\
 6 & $O(10^{-4})$ & $O(10^{-6})$ &    0.001     & $O(10^{-5})$ &   -0.025       &   -0.498     &   -0.009     \\
 7 & $O(10^{-6})$ & $O(10^{-5})$ & $O(10^{-6})$ & $O(10^{-4})$ & $O(10^{-6})$   &   -0.009     &   -0.002     \\
\hline
\end{tabular}\end{table}

\subsection{$HF^\m F^\n$ coupling}
The Higgs couplings of the dark fermion tower are given by
\begin{align}
&
\int_1^{z_L} dz\sqrt{G}
\longbar{\Psi}_F\left(-ig_AA_z\right)\gamma^5 \Psi_F\nonumber\\
&\supset -ig_Ak\int_1^{z_L} dz \frac{1}{2\sqrt{2}}\sum_{m,n}Hu_H
\left(
 f^{*(m)}_{lL}\Fbar^\m_L \gamma^5 f^{(n)}_{rR} F^\n_R
+f^{*(m)}_{rL}\Fbar^\m_L \gamma^5 f^{(n)}_{lR} F^\n_R\right)\nonumber\\
&\qquad + (L \longleftrightarrow R) \nonumber\\
&=-ig_Ak\frac{1}{\sqrt{r_{F^\m}r_{F^\n}}}\int_1^{z_L} dz\frac{1}{2\sqrt{2}}
\sin\frac{\theta_H}{2}\cos\frac{\theta_H}{2}\sum_{m,n} Hu_H\nonumber\\
\times&\bigg\{ \left(-S_L^\m(1) C_L^\m(z) C_L^\n(1) C_R^\n(z) 
+C_L^\m(1) S_L^\m(z) S_L^\n(1) S_R^\n(z) \right)F^{(m)\dagger}_LF^\n_R\nonumber\\
&+ \left(-S_L^\m(1) S_R^\m(z)C_L^\n(1) S_L^\n(z)+C_L^\m(1) C_R^\m(z)S_L^\n(1) C_L^\n(z) \right)F^{(m)\dagger}_RF^\n_L\bigg\}\nonumber\\
&=  i H \sum_{m,n} 
\left(g_{HF_L^{(m)\dagger} F^\n_R}  F_L^{(m)\dagger} F^\n_R 
- g_{HF_R^{(m)\dagger} F^\n_L} F_R^{(m)\dagger} F^\n_L \right).
\end{align} 
where
\beqn
&&\hskip -1.cm
g_{HF_L^{(m)\dagger} F^\n_R}
=g_w\frac{\sqrt{kL}}{\sqrt{r_{F^\m}r_{F^\n}}}\sin\frac{\theta_H}{2}\cos\frac{\theta_H}{2}\int_1^{z_L} dz\frac{1}{2}\frac{1}{\sqrt{z_L^2-1}} \,  z \cr
\noalign{\kern 10pt}
&&\hskip -0.cm
\times \left( S_L^\m(1) C_L^\m(z) C_L^\n(1) C_R^\n(z) 
-C_L^\m(1) S_L^\m(z) S_L^\n(1) S_R^\n(z) \right) ~, \cr
\noalign{\kern 10pt}
&&\hskip -1.cm
g_{HF_R^{(m)\dagger} F^\n_L} = 
g_w\frac{\sqrt{kL}}{\sqrt{r_{F^\m}r_{F^\n}}}\sin\frac{\theta_H}{2}\cos\frac{\theta_H}{2}\int_1^{z_L} dz\frac{1}{2}\frac{1}{\sqrt{z_L^2-1}}  \, z \cr
\noalign{\kern 10pt}
&&\hskip -0.cm
\times  \left(-S_L^\m(1) S_R^\m(z)C_L^\n(1) S_L^\n(z)
+C_L^\m(1) C_R^\m(z)S_L^\n(1) C_L^\n(z)\right) ~.
\label{HFF2}
\eeqn
One can show that 
$g_{HF_R^{(m)\dagger} F^\n_L}^* = g_{HF_R^{(m)\dagger} F^\n_L} = g_{HF_L^{(n)\dagger} F^\m_R}$.
We define
\beeq
y_{F^\m F^\n} =\frac{g_{H F_L^{(m)\dagger}F^\n_R}+g_{HF_R^{(m)\dagger}F^\n_L}}{2}  ~~,~~
\hat{y}_{F^\m F^\n}=\frac{-g_{HF_L^{(m)\dagger}F^\n_R}+g_{HF_R^{(m)\dagger}F^\n_L}}{2} ~~.
\label{HFF3}
\eneq
In particular
\beeq
y_{F^\n F^\n}  = g_w\frac{\sqrt{kL}}{\sqrt{r_{F^\m}r_{F^\n}}}
\sin\frac{\theta_H}{2}\cos\frac{\theta_H}{2} 
\frac{1}{4}\sqrt{z_L^2-1}S_L^\n(1)C_L^\n(1) ~.
\eneq
Numerical values of  $y_{F^\m F^\n}$ and $\hat{y}_{F^\m F^\n}$ are given in Table \ref{tbl:yFF} and \ref{tbl:yhFF}, respectively.

\ignore{
For $m=n$, 
\begin{align}
g_{HF_L^{(n)\dagger} F^\n_R}
=&g_w\frac{\sqrt{kL}}{\sqrt{r_{F^\m}r_{F^\n}}}\sin\frac{\theta_H}{2}\cos\frac{\theta_H}{2}\int_1^{z_L} dz\frac{1}{2}\frac{1}{\sqrt{z_L^2-1}} z\nonumber\\
&S_L^\n(1)C_L^\n(1) \left(C_L^\n(z) C_R^\n(z)-S_L^\n(z) S_R^\n(z) \right)\\
=&g_w\frac{\sqrt{kL}}{\sqrt{r_{F^\m}r_{F^\n}}}\sin\frac{\theta_H}{2}\cos\frac{\theta_H}{2}\frac{1}{4}\sqrt{z_L^2-1}S_L^\n(1)C_L^\n(1) \\
=&g_{HF_R^{(n)\dagger} F^\n_L}\nonumber\\
=&y_{HF^{(n)\dagger} F^\n}.
\end{align}
}

\begin{table}[htbp]\centering \footnotesize
\caption{$y_{F^\m F^\n}$ in the unit of $y_t\sin\frac{\theta_H}{2}$. 
The value are written by three significant figures.}\label{tbl:yFF}
\begin{tabular}{|c|ccccccc|}\hline
&1&2&3&4&5&6&7\\ \hline
 1 & -0.944 & -12.6  &  0.328 &  2.85  & -0.006 & -0.485 &  0.038 \\
 2 & -12.6  &  0.856 &  8.73  & -0.357 & -0.174 &  0.003 &  0.394 \\
 3 &  0.328 &  8.73  & -0.924 & -12.7  &  0.369 &  2.65  & -0.002 \\
 4 &  2.85  & -0.357 & -12.7  &  0.920 &  9.05  & -0.376 & -0.277 \\ 
 5 & -0.006 & -0.174 &  0.369 &  9.05  & -0.942 & -12.7  &  0.380 \\
 6 & -0.485 &  0.003 &  2.65  & -0.376 & -12.7  &  0.941 &  9.13  \\
 7 &  0.038 &  0.394 & -0.002 & -0.277 &  0.380 &  9.13  & -0.953 \\
\hline
\end{tabular}\end{table}

\begin{table}[htbp]\centering \footnotesize
\caption{$\hat{y}_{F^\m F^\n}$ in the unit of $y_t\sin\frac{\theta_H}{2}$. 
Only the value larger than $O(10^{-3})$ are shown and written by three significant figures.}\label{tbl:yhFF}
\begin{tabular}{|c|ccccccc|}\hline 
    &1&2&3&4&5&6&7\\ \hline
 1 &    0   & -6.54  &  0.117 &  1.35  & -0.017 & -0.654 &  0.013 \\
 2 &  6.54  &    0   &  1.62  & -0.094 &  0.168 &  0.005 &  0.106 \\
 3 & -0.117 & -1.62  &    0   & -1.26  &  0.066 &  0.627 & -0.002 \\
 4 & -1.35  &  0.094 &  1.26  &    0   &  0.934 & -0.057 &  0.027 \\
 5 &  0.017 & -0.168 & -0.066 & -0.934 &    0   & -0.668 &  0.044 \\
 6 &  0.654 &  0.005 & -0.627 &  0.057 &  0.668 &    0   &  0.648 \\
 7 & -0.013 & -0.106 &  0.002 & -0.027 & -0.044 & -0.648 &    0   \\
\hline\end{tabular}\end{table}

\vskip 3.cm

\providecommand{\href}[2]{#2}\begingroup\raggedright\endgroup

\end{document}